\documentclass[12pt]{article}
\catcode`\@=11
\@addtoreset{equation}{section}

\def\@seccntformat#1{\csname the#1\endcsname.~~}
\makeatother

\baselineskip=\normalbaselineskip

\usepackage{mathrsfs,amsbsy,amssymb,latexsym,amsfonts,amsmath}
\usepackage{graphicx,color}

\allowdisplaybreaks[4]
\renewcommand{\thefootnote}{\arabic{footnote}}
\def\cM{\mathcal M}
\newcommand{\tr}{\operatorname{tr}}
\def\rmd{{\mathrm d}}
\def\rmD{{\mathrm D}}
\newcommand{\Lie}{\pounds}
\newcommand{\barg}{\bar{g}}
\newcommand{\barh}{\bar{h}}
\newcommand{\barT}{\bar{T}}
\newcommand{\barK}{\bar{K}}
\newcommand{\barN}{\bar{N}}
\newcommand{\nn}{\nonumber}
\newcommand{\tc}{{\tilde{c}}}
\newcommand{\te}{{\tilde{e}}}
\newcommand{\tn}{{\tilde{n}}}
\newcommand{\tp}{{\tilde{p}}}
\newcommand{\ts}{{\tilde{s}}}
\newcommand{\tsigma}{{\tilde{\sigma}}}
\newcommand{\bk}{{\boldsymbol{k}}}
\newcommand{\bx}{{\boldsymbol{x}}}
\newcommand{\by}{{\boldsymbol{y}}}
\newcommand{\bH}{{\boldsymbol{H}}}
\newcommand{\eq}[1]{(\ref{#1})}
\newcommand{\Exp}[1]{\operatorname{e}^{#1}}
\newcommand{\ii}{{\mathrm i}}
\newcommand{\sfT}{{\mathsf T}}
\newcommand{\sfe}{{\mathsf e}}
\newcommand{\Stot}{{\hat{S}}}

\begin{document}

\begin{titlepage}
\renewcommand{\thefootnote}{\fnsymbol{footnote}}

\begin{flushright}
\parbox{3.5cm}
{KUNS-2303}
\end{flushright}

\vspace*{1.5cm}

\begin{center}
{\Large \bf Entropic formulation of\\
 relativistic continuum mechanics }%
\end{center}
\vspace{1.5cm}

\centerline{
{Masafumi Fukuma}%
\footnote{E-mail address: 
fukuma@gauge.scphys.kyoto-u.ac.jp} ~and~ 
{Yuho Sakatani}%
\footnote{E-mail address: 
yuho@gauge.scphys.kyoto-u.ac.jp}
}

\vspace{0.5cm}

\begin{center}
{\it Department of Physics, Kyoto University \\ 
Kyoto 606-8502, Japan\\}

\vspace*{1cm}

\end{center}
\vspace*{1cm}
\begin{abstract}

An entropic formulation of relativistic continuum mechanics is developed 
in the Landau-Lifshitz frame.
We introduce two spatial scales, 
one being the small scale representing the linear size of each material particle 
and the other the large scale representing the linear size of a large system 
which consists of material particles  
and is to linearly regress to the equilibrium. 
We propose a local functional which is expected to represent 
the total entropy of the larger system 
and require the entropy functional 
to be maximized in the process of linear regression. 
We show that Onsager's original idea on linear regression 
can then be realized explicitly
as current conservations with dissipative currents in the desired form. 
We demonstrate the effectiveness of this formulation 
by showing that one can treat a wide class 
of relativistic continuum materials, 
including standard relativistic viscous fluids  
and relativistic viscoelastic materials. 

\end{abstract}

\thispagestyle{empty}
\setcounter{page}{0}
\setcounter{footnote}{0}
\end{titlepage}

\newpage

\section{Introduction}

Thermodynamics is a macroscopic description of a physical system, 
taking the average of its microscopic degrees of freedom 
in both the spatial and the temporal directions. 
The spatial average is the usual coarse graining, 
while the temporal average is identified with a statistical (ensemble) average 
(see, e.g., \cite{LL_stat}). 
A global equilibrium of a given system is realized 
when the temporal average is taken sufficiently longer than 
any relaxation times of the system. 
This is characterized as the configuration which maximizes the total entropy. 
In usual observations, however, the time average is not taken for such a long time,
so the materials we often encounter are not in global equilibrium 
and are regarded as being in the process of regression to equilibrium.

We can say more about this regression to equilibrium 
if the microscopic interactions of a given system is sufficiently local 
in both the spatial and the temporal directions. 
Suppose that we divide the system into subsystems of small size 
(but still sufficiently large that thermodynamic descriptions work). 
Then, due to the locality, 
one may assume that each subsystem gets into equilibrium rapidly. 
This assumption is called the {\it local equilibrium hypothesis}. 
In the following, when writing spacetime coordinates $x=(x^\mu)$ 
we always assume that dynamical variables are already averaged 
around $x$ in both the spatial and the temporal directions 
in such a way that this local equilibrium is realized well 
with required symmetries being respected. 
Each spatial unit over which the spatial average is taken 
is called a {\it material particle}.

In the standard description of relativistic fluid mechanics,  
one adopts the formulation based on conserved currents \cite{Eckart:1940te,LL_fluid} 
(for recent developments, see \cite{review1,review2} and references therein). 
For example, when describing a simple, relativistic fluid 
in a $(D+1)$-dimensional spacetime, 
one introduces as dynamical variables the energy-momentum tensor $T^{\mu\nu}(x)$\,, 
and the particle-number current (or the charge current) $n^\mu(x)$\,;
in all, $(D+1)(D+4)/2$ degrees of freedom ($=14$ when $D=3$). 
They obey the following $D+2$ conservation laws: 
\begin{align}
  \nabla_\mu T^{\mu\nu}=0\,,\qquad \nabla_\mu n^\mu =0\,.
\label{fund_eq}
\end{align}
Meanwhile, thermodynamic arguments based on the second law of thermodynamics 
(with a few natural assumptions) tell us 
that the $(D+1)(D+4)/2$ fundamental variables can be written 
only with $D+2$ variables (and their spatial derivatives), 
and thus the conservations \eq{fund_eq} are indeed enough 
for describing the dynamics of relativistic fluids.

The method described in the previous paragraph is actually powerful 
and can be applied to a wide class of continuum materials. 
However, it is rather difficult to see 
what actually happens entropically in the process of regression to the global minimum. 
In particular, the link between Onsager's idea of linear regression \cite{Onsager:I,Onsager:II} 
and the current conservations 
has not been given in a direct manner so far. 
(An attempt was first made by Casimir \cite{Casimir}.
See also \cite{de_Groot-Mazur}.)

In this paper, we propose a framework of linear nonequilibrium thermodynamics 
which directly realizes Onsager's idea of linear regression \cite{Onsager:I,Onsager:II,Casimir} 
by introducing the explicit form of the entropy functional 
which is local and to be maximized in the process of thermalization. 
We show that linear regression to the global equilibrium 
can be naturally represented in the form of current conservations 
with dissipative currents in the desired form.

This paper is organized as follows. 
In Sec.\ 2, we give a relativistic theory of linear nonequilibrium 
thermodynamics, by closely following the idea of Onsager.  
We there introduce two spatial scales, 
$\epsilon_{\rm s}$ and $L_{\rm s}$\,.
The smaller scale $\epsilon_{\rm s}$ 
represents the linear size of each material particle, 
while the larger scale $L_{\rm s}$ represents 
that of a large system 
which consists of material particles 
and is to linearly regress to the equilibrium.  
We then propose the explicit form of the effective entropy functional 
and show that Onsager's original idea can be realized explicitly
as current conservations with dissipative currents in the desired form. 
In Sec.\ 3, we demonstrate that usual relativistic fluid mechanics 
can be reproduced within our framework. 
In Sec.\ 4, we apply the formulation to a class of continuum materials \cite{Eckart:1948}.
We propose a generally covariant generalization 
of the theory of viscoelasticity \cite{Eckart:1948,afky,afky2} 
and show that the so-called rheology equations given in \cite{afky,afky2}
can be naturally obtained in a generally covariant form. 
Section 5 is devoted to a conclusion and discussions.

\section{General theory}
\label{general_theory}

\subsection{Geometrical setup}
\label{geometrical_setup}

In order to define linear nonequilibrium thermodynamics 
in a generally covariant manner, 
we first make a few preparations. 
We consider continuum materials living 
in a $(D+1)$-dimensional Lorentzian manifold $\cM$\,.
Its local coordinates are denoted by $x^{\mu}$ 
$(\mu=0,1,\cdots,D)$ 
and the Lorentzian metric with signature $(-,+,\cdots,+)$
by $g_{\mu\nu}(x)$\,.
We take the natural unit, $\hbar = c = k_{\rm B} = 1$\,, 
throughout this paper. 
In order to develop thermodynamics, 
we first need to specify a set of timeslices (or a foliation) 
on each of which we consider spatial averages. 
In this article, we exclusively consider the foliation 
in which each timeslice is orthogonal to the energy flux 
[or the energy-momentum $(D+1)$-vector] $p_\mu$\,,
assuming that $p_\mu$ is hypersurface orthogonal. 
This choice of foliation is called the Landau-Lifshitz frame.%
We parametrize the hypersurfaces with a real parameter $t$ 
as $\Sigma_t$ $(t\in {\mathbb R})$ 
and introduce a coordinate system $x=(x^\mu)=(x^0,\,\bx)$ such that $x^0=t$\,, 
$\bx=(x^i)$ $(i=1\,,\cdots,D)$.
The unit normal $u^\mu(x)$ to the hypersurfaces 
(called the velocity field) is given by%
\begin{align}
 u^\mu(x) \equiv g^{\mu\nu}(x)\,p_\nu(x)/e(x) = p^\mu(x)/e(x)\,,
\end{align}
where $e(x)$ is the density of the proper energy (rest mass plus internal energy),
\begin{align}
  e(x) \equiv \sqrt{-g^{\mu\nu}(x)\,p_\mu(x)\,p_\nu(x)}\,.
\end{align}
Here and hereafter, indices are lowered (or raised) always with $g_{\mu\nu}$ 
(or with its inverse $g^{\mu\nu}$).
The induced metric on a $D$-dimensional hypersurface passing through $x$ 
is expressed as
\begin{align}
 h_{\mu\nu}(x)\equiv g_{\mu\nu}(x) + u_{\mu}(x)\,u_{\nu}(x)
  = g_{\mu\nu}(x) + \frac{p_\mu(x)\,p_\nu(x)}{e^2(x)}\,.
\label{hypersurface_metric}
\end{align}
We define the extrinsic curvature $K_{\mu\nu}$ of the hypersurface 
as half the Lie derivative of $h_{\mu\nu}$ 
with respect to the velocity field $u^\mu$\,: 
\begin{align}
 K_{\mu\nu}\equiv \frac{1}{2}\,\Lie_u h_{\mu\nu}
  = \frac{1}{2}\,h_\mu^{~\rho} h_\nu^{~\sigma}
  \bigl(\nabla_{\rho} u_{\sigma} + \nabla_{\sigma} u_{\rho} \bigr)\,,
\end{align}
which measures the rate of change in the induced metric 
$h_{\mu\nu}$ as material particles flow along $u^\mu$.
Note that this tensor is symmetric and orthogonal to $u^\mu$, 
$K_{\mu\nu}\,u^\nu=0$\,.

In the Arnowitt-Deser-Misner (ADM) parametrization \cite{Arnowitt:1962hi}, 
the metric and the velocity field are represented
with the lapse $N(x)$ and the shifts $N^i(x)$ $(i=1,\cdots,D)$ as
\footnote{
Throughout the present paper, we write a contravariant vector field
$v^\mu(x)$ $(\mu=0,1,\cdots,D)$ in a concise form as $v=v^\mu(x)\,\partial_\mu$\,,
regarding $\partial_\mu$ simply as a basis of a vector space
(i.e., the tangent space in a mathematical terminology).
For example, Eq.\ \eq{velocity} stands for the two equations,
$u^0=1/N$ and $u^i=N^i/N$ $(i=1,\cdots,D)$.
We denote by $v^\mu\nabla_\mu$
the covariant derivative along the vector field $v=v^\mu\,\partial_\mu$\,,
which now acts on tensor fields as a derivative operator.
}
\begin{align}
 \rmd s^2 &=g_{\mu\nu}(x)\,\rmd x^{\mu}\,\rmd x^{\nu} 
  = -N^2(x)\, \rmd t^2 
  + h_{ij}(x)\,\bigl(\rmd x^i-N^i(x)\,\rmd t\bigr)\,
   \bigl(\rmd x^j-N^j(x)\,\rmd t \bigr)\,,
\label{ADM_decomp}\\
 u &= u^\mu(x)\,\partial_\mu = \frac{1}{N(x)}\,\partial_t 
  + \frac{N^i(x)}{N(x)}\,\partial_i 
 \quad \bigl( \Leftrightarrow~ u_\mu(x)\,\rmd x^\mu = -N(x)\,\rmd t\bigr)\,.
\label{velocity}
\end{align}

With a given foliation, we still have the symmetry of 
foliation preserving diffeomorphisms  
that give rise to transformations only among the points on each timeslice. 
Using this residual gauge symmetry 
we can impose the {\it synchronized gauge}, $N^i(x)\equiv 0$\,, 
so that the background metric and the velocity field become
\begin{align}
 \rmd s^2 &= g_{\mu\nu}(x)\,\rmd x^\mu\,\rmd x^\nu
  \equiv -N^2(x)\,\rmd t^2 + h_{ij}(x)\,\rmd x^i\,\rmd x^j\,,\\
 &(h_{\mu\nu}) = \left(\begin{matrix}
   0 & 0 \cr 0 & h_{ij} \end{matrix}\right)\,,\quad
  (h_\mu^{~\nu}) = \left(\begin{matrix}
   0 & 0 \cr 0 & \delta_i^j \end{matrix}\right)\,,\\
 u&=u^\mu(x)\,\partial_\mu 
  = \frac{1}{N(x)}\,\frac{\partial}{\partial t} = \frac{\partial}{\partial \tau}\,,
\end{align}
where $\tau$ is the local proper time defined by $\rmd \tau=N\,\rmd t$\,.
The volume element on the hypersurface through spacetime point $x$ 
is given by the $D$-form $\sqrt{h}\,\rmd^D\bx
  = \sqrt{\det(h_{ij})}\,\rmd^D\bx$\,.
The volume element of the total $(D+1)$-dimensional manifold 
is given by $\sqrt{-g}\,\rmd^{D+1}x=N\sqrt{h}\,\rmd t\,\rmd^D \bx$\,.
Note that even after taking the synchronized gauge, 
there remains a residual gauge symmetry 
of reparametrizing $t$ that labels the timeslices.

For generic coordinates (i.e., not necessarily the synchronized coordinates), 
we introduce the {\it time derivative} $\rmD/\rmD t$ 
as the operation that satisfies the following conditions:
\begin{align}
 \mbox{(i) }\frac{\rmD}{\rmD t}\,t = 1\,,\quad
 \mbox{(ii) }\frac{\rmD}{\rmD t}\,u^\mu = 0\,,\quad
 \mbox{(iii) }\mbox{Leibniz rule}.
\end{align}
The first condition ensures that $\Delta t\,(\rmD/\rmD t)$ certainly measures 
the difference between a quantity on $\Sigma_{t+\Delta t}$ 
and that on $\Sigma_t$\,.
The second condition ensures that if a system of small spatial region
can be regarded as being static at time $t$, 
so can be the system at time $t+\Delta t$\,.
An obvious solution is given by
the Lie derivative along the velocity $u=u^\mu\partial_\mu$ 
multiplied with the lapse $N$\,:
\begin{align}
 \frac{\rmD}{\rmD t} \equiv N \Lie_u\,.
\end{align}
This also satisfies the condition (iii).
In the following, the time derivative $\rmD/\rmD t$ 
will be often abbreviated as the dot.

\subsection{Thermodynamic variables}
\label{thermodynamic_variables}

\begin{figure}[htbp]
\vspace{3ex}
\begin{center}
\includegraphics[scale=0.6]{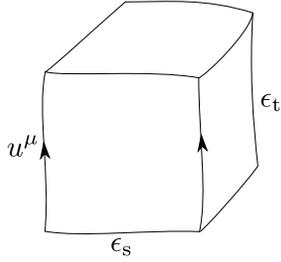}
\begin{quote}
\vspace{-5mm}
\caption{Local thermodynamic equilibrium.\label{fig:local_thermodynamic_equilibrium}}
\end{quote}
\end{center}
\vspace{-6ex}
\end{figure}

As was discussed in Introduction, 
around each spacetime point $x=(x^0=t,\bx)$ on timeslice $\Sigma_t$\,, 
we make spatial and temporal (or ensemble) averages 
over $(D+1)$-dimensional regions 
whose linear sizes we denote by 
$\epsilon_{\rm s}$ and $\epsilon_{\rm t}$\,, respectively 
(see Fig.\ \ref{fig:local_thermodynamic_equilibrium}).
We assume that $\epsilon_{\rm s}$ and $\epsilon_{\rm t}$ 
are both much smaller than the curvature radius 
of the background metric $g_{\mu\nu}$\,, 
so that we can take a local inertial frame at each spacetime point $x$ 
which can be regarded as being flat at least within the extension 
of $(\epsilon_{\rm s},\,\epsilon_{\rm t})$\,.
This implies that thermodynamic properties of each material particle 
do not depend on the curvature of the metric
when discussing thermodynamics in each material particle.
Since the affine connection is not covariant, 
our assumption means that the covariant local entropy at $x$ 
depends only on the local value of the metric, $g_{\mu\nu}(x)$\,.

We assume that the local thermodynamic properties 
of the material particle at $x$ (already in its local equilibrium) 
are specified by the set of local quantities $\bigl(b^A(x)$,\,$c^I(x)$,\,
$d^P(x)\bigr)$\,.
Here $c^I(x)$ denote the densities of the existing additive conserved quantities $C^I$\,.
$b^A(x)$ denote the ``intrinsic'' intensive variables 
possessed by each material particle (such as strains), 
and $d^P(x)$ denote the rest ``external'' intensive variables 
which further need to be introduced 
to characterize each subsystem thermodynamically 
(such as the background electromagnetic or gravitational fields). 
In this paper, we distinguish density quantities from other intensive quantities, 
and construct, by multiplying them with the spatial volume element $\sqrt{h}$\,, 
new quantities which are spatial densities on each timeslice.
For example, the entropy density $s$ and the densities $c^I$ of conserved charges 
are density quantities, 
and for them we construct the following spatial densities:
\begin{align}
 \ts\equiv \sqrt{h}\,s\,,\qquad \tc^I\equiv \sqrt{h}\,c^I\,.
 \label{tilde_notation}
\end{align}
The local equilibrium hypothesis implies that 
the local entropy $\ts(x)$ is already maximized 
at each spacetime point $x$ 
and is given as a function of the above local variables; 
$\ts(x)=\ts\bigl(b^A(x),\,\tc^I(x),\,d^P(x)\bigr)$\,.
This functional relation is sometimes called 
the {\it fundamental relation in the entropy representation}.

In the synchronized gauge, 
due to the relation $\partial/\partial t=N(x)\,\partial/\partial\tau$\,,  
the proper energy density $e(x)$ measured with the proper time $\tau$ 
is related to the energy density $\sfe(x)$ measured with time $t$ as 
\begin{align}
 \sfe(x) = N(x)\,e(x)\,.
\end{align}
Note that $\sfe(x)$ includes the gravitational potential through the factor $N(x)$\,.
Accordingly, 
the local temperature $T$ measured with $\tau$ 
$\bigl(T\equiv (\partial \ts/ \partial \te)^{-1}\bigr)$
is related to the temperature $\sfT$ measured with $t$ 
$\bigl(\sfT \equiv (\partial \ts/ \partial \tilde{\sfe})^{-1}\bigr)$ 
through the following Tolman law:
\begin{align}
 \sfT(x) = N(x)\,T(x)\,.
\label{Tolman}
\end{align}

\subsection{Entropy functional and the current conservations}
\label{entropy_functional}

A local thermodynamic equilibrium is realized 
only in each material particle of spatial size $\epsilon_{\rm s}$ 
averaged for a period of time, $\epsilon_{\rm t}$\,. 
If we observe a material at spacetime scales 
larger than $(\epsilon_{\rm s},\,\epsilon_{\rm t})$\,, 
we need to take into account the effect 
that the material particles communicate with each other 
by exchanging conserved quantities 
(e.g., the energy-momentum and the particle number). 
The second law of thermodynamics tells us that, 
if boundary effects can be neglected, 
this should proceed such that the total entropy of the larger region gets increased.

\begin{figure}[htbp]
\vspace{3ex}
\begin{center}
\includegraphics[scale=0.6]{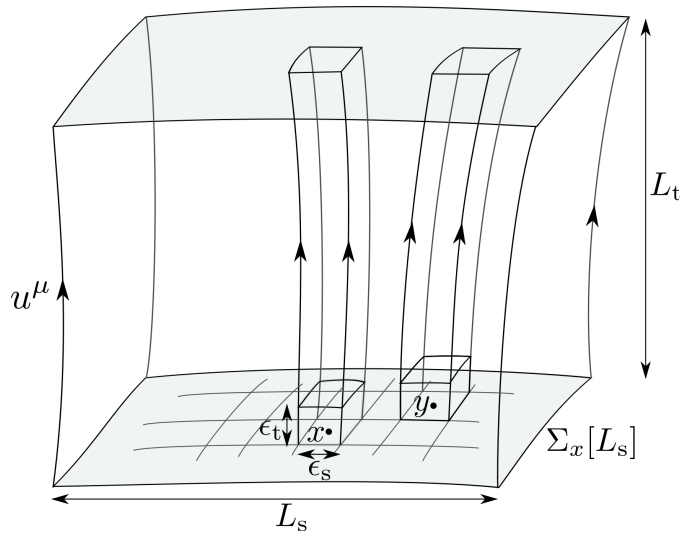}
\begin{quote}
\vspace{-5mm}
\caption{Time evolution of material particles 
in the large region $\Sigma_x[L_{\rm s}]$\,.
\label{fig:large_region}}
\end{quote}
\end{center}
\vspace{-6ex}
\end{figure}

In order to describe such dynamics mathematically,  
we first introduce another spacetime scale $(L_{\rm s},\,L_{\rm t})$ 
which is much larger than $(\epsilon_{\rm s},\,\epsilon_{\rm t})$ 
and assign to each spacetime point $x=(x^0=t,\bx)$ 
on the timeslice $\Sigma_t$ 
a spatial region $\Sigma_x[L_{\rm s}]$ of linear size $L_{\rm s}$ 
(see Fig.\ \ref{fig:large_region}).
We then consider the total entropy of the region $\Sigma_x[L_{\rm s}]$:
\begin{align}
 \Stot(t;\,\Sigma_x[L_{\rm s}]) 
  &\equiv
   \int_{\Sigma_x[L_{\rm s}]} \rmd^D\by\,
             \ts\bigl(b^A(t,\by),\,\tc^I(t,\by),\,d^P(t,\by)\bigr) \,. 
\end{align}
The irreversible motion of intrinsic variables 
$a^r(x)\equiv \bigl(b^A(x),\,\tc^I(x)\bigr)$ at $x$ 
will proceed toward 
the equilibrium state of the region $\Sigma_x[L_{\rm s}]$\,.
Due to the condition $L_{\rm s} \gg \epsilon_{\rm s}$\,, 
we can assume that the influence from the surroundings of 
the region $\Sigma_x[L_{\rm s}]$ is not relevant 
to the dynamics of $a^r(x)$ because $x$ is well inside the region.
An equilibrium state of the region $\Sigma_x[L_{\rm s}]$ 
will be realized when the observation is made 
for a long period of time, $L_{\rm t}$\,, 
and can be characterized by the condition%
\begin{align}
 \frac{\delta \Stot(t;\,\Sigma_x[L_{\rm s}]) }{\delta a^r(x)}=0\,.
\label{equilib_value}
\end{align}
Note that the functional derivative is taken 
only with respect to a spatial, $D$-dimensional unit in the functional.
We denote the values of $a^r(x)$ at the equilibrium by 
$a^r_0(x;\,L_{\rm s})
 \equiv \bigl(b^A_0(x;\,L_{\rm s}),\,\tc^I_0(x;\,L_{\rm s})\bigr)$\,,
and will call the procedure to obtain $a^r_0(x;\,L_{\rm s})$ from $a^r(x)$ 
the {\it dynamical block-spin transformation}. 
One here should note that, 
since $\tc^I(t,\by)$ are the densities of conserved quantities,
the variations \eq{equilib_value} with respect to $\tc^I$-type variables 
should be taken with total charges kept fixed at prescribed values: 
\begin{align}
 \int_{\Sigma_x[L_{\rm s}]}\rmd^D\by \,\tc^I(t,\by)
  \equiv C^I\bigl(\Sigma_x[L_{\rm s}]\bigr) \,.
\label{conservation_law}
\end{align}
A simple analysis using the Lagrange multipliers shows 
that the condition of global equilibrium is expressed locally as
\begin{align}
 \frac{\partial \ts}{\partial b^A}(x)=0 \qquad \text{and}\qquad
  h_\mu^{~\nu}(x)\,\nabla_\nu \beta_I(x) = 0 \,,
\label{equilibrium_condition}
\end{align}
where $\beta_I$ is the thermodynamic variable conjugate to $\tc^I$ 
that is defined by
\begin{align}
 \beta_I(x) \equiv \frac{\partial \ts}{\partial \tc^I}(x) \,.
\end{align}

We here make a few comments. 
First, the procedure to define the equilibrium values 
$a^r_0(x;\,L_{\rm s})$ at point $x$ is carried out 
over its own region $\Sigma_x[L_{\rm s}]$\,; 
for a point $y$ different from $x$\,, $a^r_0(y;\,L_{\rm s})$ 
should be obtained by solving the equation
$\delta\Stot(t;\,\Sigma_y[L_{\rm s}]) /\delta a^r(y) = 0$\,.
However, when $y$ is well inside the region $\Sigma_x[L_{\rm s}]$\,,
in the approximation of linear regression 
the value of $a^r_0(y;\,L_{\rm s})$ can be regarded as the same 
with the value $a^r_0(y;\Sigma_x[L_{\rm s}])$ that is obtained by solving the equation 
$\delta\Stot(t;\,\Sigma_x[L_{\rm s}]) /\delta a^r(y) = 0$\,.

\begin{figure}[htbp]
\vspace{3ex}
\begin{center}
\includegraphics[scale=0.35]{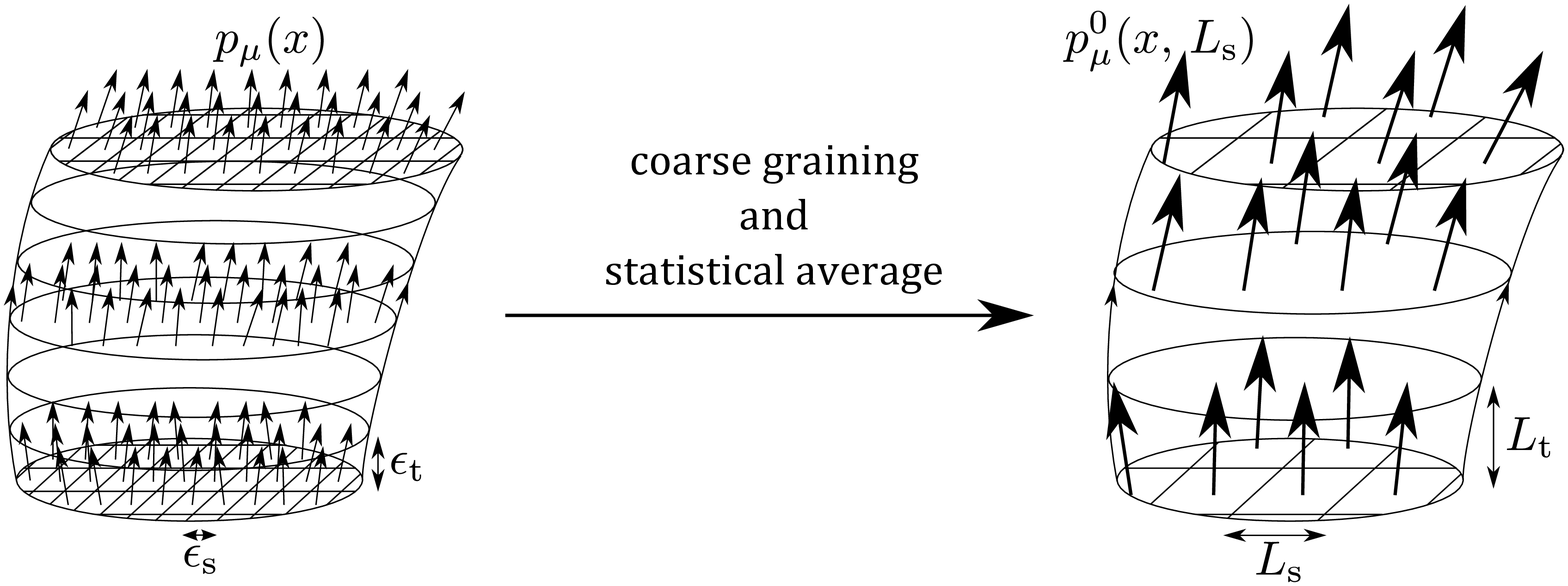}
\begin{quote}
\vspace{-5mm}
\caption{Dynamical block-spin transformation 
from the small-scale variable $p_\mu(x)$ (left)
to the large-scale variable $p_\mu^0(x,\,L_{\rm s})$ (right).
\label{fig:large_small}}
\end{quote}
\end{center}
\vspace{-6ex}
\end{figure}

The second comment is about the spatial scale $L_{\rm s}$\,. 
We take $L_{\rm s}$ sufficiently larger than $\epsilon_{\rm s}$ 
such that $a^r_0(y;\,L_{\rm s})$ themselves can be treated 
as thermodynamic variables%
\footnote{
Their fluctuations can be roughly estimated 
to be $\Delta a^r_0/\langle a^r_0\rangle\sim(\epsilon_{\rm s}/L_{\rm s})^{D/2}$\,.
} 
as well as that the boundary effects can be safely neglected.
At the same time, 
we also take the spacetime scale $(L_{\rm s},\,L_{\rm t})$ 
not too much larger than 
$(\epsilon_{\rm s},\,\epsilon_{\rm t})$ 
such that the values of $a^r(y)$ $\bigl(y=(t,\by)\in\Sigma_x[L_{\rm s}]\bigr)$ 
do not differ significantly from those of $a^r_0(x;\,L_{\rm s})$\,.
The latter condition ensures that 
the fluctuation of $a^r(y)$ around $a^r_0(y;\,L_{\rm s})$ 
can be well approximated by the Gaussian distribution.%
\footnote{
We also require that $(L_{\rm s},\,L_{\rm t})$ be 
much smaller than the typical scales 
over which the dynamics may change substantially 
(such as the typical size of a continuum material 
or the typical scale of the change in background fields). 
This ensures that the dependence of $a^r_0(x;\,L_{\rm s})$ on 
the scale $L_{\rm s}$ is negligibly mild.
} 
So long as $L_{\rm s}$ is chosen in this way,
the time evolution of the local variables $a^r(x)$ at time scale $\epsilon_{\rm t}$
can be analyzed elaborately by decomposing $a^r(x)$ 
into the large-scale variable $a_0^r(x)$ 
and the small-scale variable $(a-a_0)^r(x)$\,,
\begin{align}
 a^r(x) \equiv a_0^r(x) + (a-a_0)^r(x)\,.
\label{large-small}
\end{align}
In fact, the variables $a_0^r(x)$ 
evolve with scale $(L_{\rm s},\,L_{\rm t})$\,, 
and their variations at time scale $\epsilon_{\rm t}$ 
can actually be regarded as being negligibly small (see Fig.\ \ref{fig:large_small})\,.
In contrast, the variables $(a-a_0)^r(x)$ represent the Gaussian fluctuations 
in the region $\Sigma_x[L_{\rm s}]$\,, 
and their evolutions can be analyzed with the dynamics 
of spacetime scale $(\epsilon_{\rm s},\,\epsilon_{\rm t})$ 
in the linear approximation.

We now recall Onsager's discussion on the linear regression 
of nonequilibrium system \cite{Onsager:I,Onsager:II,Casimir,LL_stat}. 
We first decompose the time derivative of $a^r(x)$\,,
$\dot{a}^r(x)\equiv \rmD a^r(x)/\rmD t$\,, 
into the reversible and irreversible parts:
\begin{align}
 \dot{a}^r(x) \equiv [\dot{a}^r(x)]_{\rm rev} + [\dot{a}^r(x)]_{\rm irr}\,.
\end{align}
With the decomposition of $a^r(x)$ itself [Eq.\ \eq{large-small}], 
the reversible part can be written as
\begin{align}
 [\dot{a}^r(x)]_{\rm rev}=\dot{a}_0^r(x) 
  + \bigl[(\dot{a}-\dot{a}_0)^r(x)\bigr]_{\rm rev}\,,
\end{align}
where the first term describes the reversible, isentropic motion 
of the large-scale variable $a^r_0(x)$\,, 
and the second represents the part of $(\dot{a}-\dot{a}_0)^r(x)$ 
that does not affect the increase of the total entropy 
$\Stot(t;\Sigma_x[L_{\rm s}])$\,.
Since in the current discussions 
$a^r(x)$ 
deviate from the equilibrium values 
$a_0^r(x)$ only slightly, 
we can assume that the irreversible motion of the small-scale variables 
obeys the linear regression law,%
\begin{align}
 [\dot{a}^r(x)]_{\rm irr} = \bigl[(\dot{a}-\dot{a}_0)^r(x)\bigr]_{\rm irr}
  = -\,\gamma^{\,r}_{~s}\,(a-a_0)^s(x)\,.
\label{linear_regression}
\end{align}
Here the matrix $\gamma=(\gamma^{\,r}_{~s})$ 
has only positive eigenvalues, 
and we have used the relation 
$\dot{a}^r_0(x)=[\dot{a}^r_0(x)]_{\rm rev}$\,, 
that is, $[\dot{a}^r_0(x)]_{\rm irr}=0$\,.
We then expand the total entropy 
$\Stot(t;\,\Sigma_x[L_{\rm s}])$
around $a^r_0(x)$ to second order 
and write the difference between the total entropy for the configuration $a^r(x)$\,,
$\Stot(t;\,\Sigma_x[L_{\rm s}])
 =\int_{\Sigma_x[L_{\rm s}]} \rmd^D\by\,
 \ts(b^A(t,\by),\,\tc^I(t,\by),\,d^P(t,\by))$\,, 
and that at the equilibrium $a^r_0(x)$\,, 
$\Stot_0(t;\,\Sigma_x[L_{\rm s}])\equiv 
 \int_{\Sigma_x[L_{\rm s}]} \rmd^D\by\, \,
 \ts(b_0^A(t,\by),\,\tc_0^I(t,\by),\,d^P(t,\by))$\,,
as
\begin{align}
 \Delta\Stot(t;\,\Sigma_x[L_{\rm s}])
 &\equiv \Stot(t;\,\Sigma_x[L_{\rm s}])
  -\Stot_0(t;\,\Sigma_x[L_{\rm s}])\nn\\ 
 &= - \,\frac{1}{2} \,\int_{\Sigma_x[L_{\rm s}]} 
                    \rmd^D\by\, \, (a-a_0)^r\,\beta_{rs}\,(a-a_0)^s\,.
\label{entropy_functional_expansion}
\end{align}
Here $\beta_{rs}$ is a symmetric, positive semidefinite operator 
(which may include spatial derivative operators). 
The thermodynamic forces are then found to be%
\begin{align}
 f_r(x) \equiv 
  \frac{\delta\,\Delta\Stot(t;\,\Sigma_x[L_{\rm s}])}{\delta a^r(x)}
  = -\,\beta_{rs}\,(a-a_0)^s(x)\,,
\end{align}
and thus, substituting them into Eq.\ \eq{linear_regression}, 
we obtain the equations of linear regression:
\begin{align}
 [\dot{a}^r(x)]_{\rm irr} = L^{rs}\,f_s(x)
  = \,L^{rs}\,
   \frac{\delta\,\Delta\Stot(t;\,\Sigma_x[L_{\rm s}])}{\delta a^s(x)}
 \quad \bigl(L^{rs}\equiv (\gamma\,\beta^{-1})^{rs}\bigr)\,.
\label{Onsager_eq}
\end{align}

The so-called phenomenological coefficients $L^{rs}$ 
can be shown to satisfy 
Onsager's reciprocal relation \cite{Onsager:I,Onsager:II,Casimir}
\begin{align}
 L^{rs} = (-1)^{|a^r| + |a^s|}\,L^{sr}\,,
\end{align}
where the index $|a^r|$ expresses how the variables transform 
under time reversal, 
$a^r(x)\to (-1)^{|a^r|}\,a^r(x)$\,.%
\footnote{
When the system is subject to external fields $\bH$ 
which change the sign under time reversal (like magnetic field), 
the reciprocal relation is expressed as
$L^{rs}(\bH) = (-1)^{|a^r| + |a^s|}\,L^{sr}(-\bH)$\,.
} 
The Curie principle says that $L^{rs}$ can be block diagonalized 
with respect to the transformation properties of the indices $(r,s)$ 
under spatial rotations and the parity transformation \cite{de_Groot-Mazur}, 
that is, under the subgroup $\mathrm{O}(D)$ of the local Lorentz group $\mathrm{O}(D,1)$ 
in local inertial frames. 
For example, when $a^r$ constitute a contravariant vector, $(a^r)\equiv(a^\mu)$\,, 
the equations of linear regression 
should be set for each of the normal and tangential components 
to the timeslice through $x$\,:
\begin{align}
  [\dot{a}(x)]_{{\rm irr}\,\bot}^{\mu}(x) 
  &= L_\bot^{\mu\nu}\,\biggl[\frac{\delta\,\Delta\Stot}
  {\delta a^\nu(x)}\biggr]_{\mbox{\raisebox{4pt}{\scriptsize{$\bot$}}}}\,,\\
 [\dot{a}(x)]_{{\rm irr}\,\|}^{\mu}(x) 
  &= L_\|^{\mu\nu}\,\biggl[\frac{\delta\,\Delta\Stot}
  {\delta a^\nu(x)}\biggr]_{\mbox{\raisebox{4pt}{\scriptsize{$\|$}}}}\,,
\end{align}
where for a contravariant vector $v^\mu$ 
we define $v_\bot^\mu\equiv (-u^\mu u_\nu)\,v^\nu$ 
and $v_\|^\mu\equiv h^\mu_{~\nu}\,v^\nu$ 
(and similarly for covariant vectors).%
Covariance and positivity further imposes the condition 
that $L_\bot^{\mu\nu}$ and $L_\|^{\mu\nu}$ should be expressed as
$L_\bot^{\mu\nu} = L_\bot\,u^\mu u^\nu$ $(L_\bot>0)$ and 
$L_\|^{\mu\nu} = L_\|\,h^{\mu\nu}$ $(L_\|>0)$\,, respectively.

In Eq.\ \eq{entropy_functional_expansion}, 
the total entropy $\Stot(t;\,\Sigma_x[L_{\rm s}])$ 
is expanded around the equilibrium states $a^r_0(y)$ 
($y\in\Sigma_x[L_{\rm s}]$) as a quadratic form in $(a-a_0)^r(y)$\,.
However, due to the equilibrium condition \eq{equilibrium_condition},
it should be more efficient to express it as a quadratic form in
$(b-b_0)^A(y)$ and $h_\mu^{~\nu}\,\nabla_\nu \beta_I(y)$\,, 
because we then can ignore the subtleness 
existing in the constrained variations with respect to $\tc^I$-type variables. 
This consideration leads us to propose 
that the total entropy can be effectively expressed 
by the following local functional 
(to be called the {\it entropy functional}):
\begin{align}
 \Stot(t;&\,\Sigma_x[L_{\rm s}]) \nn\\
 \equiv &\, \Stot_0(t;\,\Sigma_x[L_{\rm s}])
    - \,\frac{1}{2} \,\int_{\Sigma_x[L_{\rm s}]} \rmd^D \by \,N \sqrt{h}\, 
 \begin{pmatrix} (b-b_0)^A & 
  \nabla_\mu \beta_I
 \end{pmatrix} 
  \begin{pmatrix}
  \ell_{AB}  & \ell_A^{\,\nu J}\cr
  \ell_B^{\,\mu I} & \ell^{\,\mu I,\, \nu J}
  \end{pmatrix}
 \begin{pmatrix} (b-b_0)^B \cr 
  \nabla_\nu \beta_J
 \end{pmatrix} \nn\\
 &  \bigl( \ell_A^{\,\mu I}\,u_\mu = 0 = \ell^{\,\mu I,\, \nu J}\,u_\nu \bigr)\,.
\label{total_entropy}
\end{align}
Here the coefficient $\Bigl(\begin{smallmatrix}
  \ell_{AB}  & \ell_A^{\,\nu J}\cr
  \ell_B^{\,\mu I} & \ell^{\,\mu I,\, \nu J}
  \end{smallmatrix}\Bigr)$ 
is a symmetric, positive semidefinite matrix.
We have extracted the factor $N$ from the coefficient matrices 
for later convenience (recall that $N\sqrt{h}=\sqrt{-g}$).
The matrix elements in principle could be calculated from 
the fundamental relation, $\ts=\ts(b^A,\,\tc^I,\,d^P)$\,,
once the foliation is fixed 
[see the Appendix \ref{entropy_derivation} for the derivation of \eq{total_entropy} 
for simple cases]. 
However, as we see below, their explicit forms need not be specified 
for the following discussions.

With the entropy functional \eq{total_entropy},
the thermodynamic forces in the equations of linear regression 
Eq.\ \eq{Onsager_eq} 
are calculated as
\begin{align}
 f_A= \frac{\delta\Delta\Stot}{\delta b^A}
  &= -\,\sqrt{-g}\,\bigl[\ell_{AB}\,(b -b_0)^B
    +\ell^{\mu I}_A\,\nabla_\mu \beta_I 
  -s_{AI} \,\nabla_\mu \,\bigl(
  \ell^{\mu I}_B\,(b-b_0)^B 
  + \ell^{\mu I,\,\nu J}\, \nabla_\nu \beta_J \bigr)\bigr]\nn\\
 &\simeq -\,\sqrt{-g}\,\bigl[\ell_{AB}\,(b-b_0)^B
    +\ell^{\mu I}_A\,\nabla_\mu \beta_I  \bigr]\,,
\label{thermodynamic_force_b}
\\
 f_I = \frac{\delta\Delta\Stot}{\delta \tc^I} 
  &=  \sqrt{-g}\,(\partial^2\ts/\partial\tc^I\partial\tc^J)\, 
  \nabla_\mu\bigl[\ell_{B}^{\mu J}\,(b-b_0)^B
  + \ell^{\mu J,\nu K}\, \nabla_\nu\beta_K\bigr]\nn\\
 &\simeq N s^0_{IJ}\, \nabla_\mu\bigl[\ell_{B}^{\mu J}\,(b-b_0)^B
  + \ell^{\mu J,\nu K}\, \nabla_\nu\beta_K\bigr]\,,
\label{thermodynamic_force_c}
\end{align}
where $s_{AI}\equiv \partial^2\ts/\partial b^A\partial\tc^I$\,,\ 
$s_{IJ}\equiv \sqrt{h}\,\partial^2\ts/\partial\tc^I\partial\tc^J$
and $s^0_{IJ}$ is the value of $s_{IJ}$ at the equilibrium.
In obtaining the last line of Eq.\ \eq{thermodynamic_force_b}
we have neglected higher-order terms in the derivative expansion. 
Then the 
irreversible 
evolution of $(b^A,\,\tc^I)$ is given by%
\begin{align}
 \bigl[\dot{b}\bigr]^A_{\rm irr}     &= \frac{1}{\sqrt{h}}\,L^{AB}\,f_B + L^{AI}\,f_I \,,\\
 \bigl[\dot{\tc}\bigr]^I_{\rm irr} &= L^{IA}\,f_A + \sqrt{h}\,L^{IJ}\,f_J \,,
\end{align}
where we have multiplied $(\sqrt{h})^{\pm 1}$ to make $L^{rs}$ tensors 
(not tensor densities).
By further assuming that $L^{IA}=L^{AI}=0$ 
and by using Eqs.\ \eq{thermodynamic_force_b} and \eq{thermodynamic_force_c}, 
the above equations can be rewritten as
\begin{align}
 \dot{b}^A -\bigl[\dot{b}\bigr]^A_{\rm rev} =
  \bigl[\dot{b}\bigr]^A_{\rm irr} &= -\,N\,L^{AB}\,\bigl[\,\ell_{BC}\,(b-b_0)^C
                        + \ell^{\mu I}_B\,\nabla_\mu \beta_I\,\bigr] \,,
\label{regression_b}\\
 \dot{\tc}^I -\bigl[\dot{\tc}\bigr]^I_{\rm rev} = 
  \bigl[\dot{\tc}\bigr]^I_{\rm irr} &= \sqrt{-g}\,L^{IJ}\, s_{JK}^0\, 
  \nabla_\mu \bigl[\,\ell_B^{\mu K}\,(b-b_0)^B
                   + \ell^{\mu K,\nu L}\, \nabla_\nu \beta_L\,\bigr] \,.
\label{regression_c}
\end{align}
The coefficient $L^{IJ}\,s^0_{JK}$ can be regarded as being constant 
in our derivative expansion, 
and, as we see in the next section, 
for the cases of interest 
the quantities $\dot{\tilde{c}}^I=N \Lie_u(\sqrt{h}\,c^I)$ can be always written 
as $\sqrt{-g}\,\nabla_\mu(c^I u^\mu)$\,.
We thus can rewrite Eq.\ \eq{regression_c} 
in the form of current conservations:
\begin{align}
 \nabla_\mu J^{I \mu}= 0\,.
\end{align}
Here the current $J^{I\mu}$ consists of three parts:
\begin{align}
 J^{I\mu} &= c^I u^\mu + J_{\rm (r)}^{I\mu} + J_{\rm (d)}^{I\mu}\,.
\end{align}
The first is the convective part, 
and the third is the dissipative part defined by
\begin{align}
 J_{\rm (d)}^{I\mu}\equiv
  -L^{IJ}\, s_{JK}^0\, 
  \bigl[\,\ell_B^{\mu K}\,(b-b_0)^B
  + \ell^{\mu K,\nu L}\, \nabla_\nu \beta_L\,\bigr]\,.
\end{align}
The second comes from 
the equation $\bigl[\dot{\tc}\bigr]^I_{\rm rev}
=-\sqrt{-g}\,\nabla_\mu J^{I\mu}_{\rm (r)}$\,,
which should be derived from another theory 
describing reversible, isentropic processes. 
For example, for simple viscous fluids, 
such theory should be that of ideal fluids.

When there are only $c^I$-type variables which are all scalar functions, 
the coefficients $\ell^{\mu I,\,\nu J}$ should take the form
$\ell^{\mu I,\,\nu J}=-\ell\,h^{\mu\nu}\,(s_0^{-1})^{IJ}$ ($\ell\geq 0$)
due to the positivity and the covariance, 
and the dissipative current is expressed as
\begin{align}
 J_{\rm (d)}^{I\mu}\equiv
  \ell\,L^{IJ}\,h^{\mu\nu}\,\nabla_\nu \beta_J
  =\ell\,L^{IJ}\,h^{\mu\nu}\,\nabla_\nu \Bigl( 
  \frac{\partial \ts}{\partial \tc^J}\Bigr)\,.
\end{align}

The second law of thermodynamics says 
that $\Stot(t;\,\Sigma_x[L_{\rm s}])$ 
is a monotonically increasing function of time $t$\,.
Indeed, the entropy production rate $P(t;\,\Sigma_x[L_{\rm s}])$ 
in the region $\Sigma_x[L_{\rm s}]$
is always positive for time scale $\epsilon_{\rm t}$  
due to the positivity of the phenomenological coefficients:%
\begin{align}
 P(t;\,\Sigma_x[L_{\rm s}])&\equiv 
   \frac{\rmD }{\rmD t}\Delta S(t;\,\Sigma_x[L_{\rm s}])
  = \int_{\Sigma_x[L_{\rm s}]} \rmd^D\by\,
   \frac{\delta\Delta \Stot}{\delta a^r} \,
   \dot{a}^r\nn\\
 &= \int_{\Sigma_x[L_{\rm s}]} \rmd^D\by\,
   \frac{\delta \Delta\Stot}{\delta a^r} \,
   [\dot{a}^r]_{\rm irr}
  = \int_{\Sigma_x[L_{\rm s}]} \rmd^D\by\, L^{rs}\,
   \frac{\delta\Delta\Stot}{\delta a^r}\,
   \frac{\delta\Delta\Stot}{\delta a^s}
  \geq 0\,.
\label{entropy_production}
\end{align}
Here we have neglected the contributions from $\dot{d}^P$ 
because they would be of higher orders.
We also have used the identity $\int_{\Sigma_x[L_{\rm s}]} \rmd^D\by
(\delta\Delta\Stot/\delta a^r)\,[\dot{a}^r]_{\rm rev}=0$\,.
One can further show that the entropy production is 
a monotonically decreasing function 
when $[\dot{a}^r]_{\rm rev}=0$\,; 
$\dot{P}(t;\,\Sigma_x[L_{\rm s}]) \leq 0$\,.

\section{Relativistic fluid mechanics}
\label{relativistic_fluid_mecanics}

\subsection{Relativistic fluids in the Landau-Lifshitz frame}
\label{L-L_frame}

We define relativistic fluids as continuum materials 
whose thermodynamic properties can be characterized 
only by the local energy-momentum $\tp_\alpha=\sqrt{h}\,p_\alpha$ 
($\alpha=0,1,\cdots,D$)\,,
the local particle number $\tn=\sqrt{h}\,n$\,, 
and the background metric $g_{\mu\nu}$\,.

To relativistic fluids, the formalism in the previous section 
can be applied with the following identifications:
\begin{center}
\begin{tabular}{|c|c||c|c||c|c|}
\hline
 $b^A$ &  $(\partial\ts/\partial b^A)$ &
 $\tc^I$   & $\beta_I$ &  $d^P$ &  $(\partial\ts/\partial d^P)$ \\ \hline\hline
 nothing & nothing &
 $\tp_\alpha$    &    $ - u^\alpha/T$ & $g_{\mu\nu}$ 
  & $\sqrt{h}\,T^{\mu\nu}_{\rm (q)}/2T$ \\
 & &
 $\tn$ & $-\mu/T$ & & \\
\hline
\end{tabular}
\end{center}
We here explain the entities in the list. 
We shall assume that $\ts$ depends on $\tp_\alpha$  
only through the local energy 
$\te(\tp_\alpha,g_{\mu\nu}) = \sqrt{-g^{\mu\nu}\,\tp_\mu\,\tp_\nu}$ 
so that the fundamental relation is expressed in the following form: 
\begin{align}
 \ts(\tp_\alpha, \tn, g_{\mu\nu})
  = \tsigma\bigl( \te(\tp_\alpha, g_{\mu\nu}),\tn,g_{\mu\nu}\bigr)\,.
\label{fundamental_relation}
\end{align}
Then, by introducing the temperature $T$\,, the chemical potential $\mu$ 
and the quasiconservative stress $\tau_{\rm (q)}^{\mu\nu}$ as
\begin{align}
 \frac{\partial\tsigma(\te,\tn,g_{\mu\nu})}{\partial\te}\equiv\frac{1}{T}\,,\quad
  \frac{\partial\tsigma(\te,\tn,g_{\mu\nu})}{\partial\tn}\equiv -\frac{\mu}{T}\,,\quad
  \frac{\partial\tsigma(\te,\tn,g_{\mu\nu})}{\partial g_{\mu\nu}}
  \equiv\frac{\sqrt{h}}{2T}\,\tau_{\rm (q)}^{\mu\nu}\,,
\end{align}
and by using the identities
\begin{align}
 \frac{\partial\te}{\partial\tp_\alpha} = -\frac{\tp^\alpha}{\te}=-u^\alpha\,,\quad
  \frac{\partial\te}{\partial g_{\mu\nu}}
  = \frac{\tp^\mu \tp^\nu}{2\,\te} = \frac{\te\,u^\mu u^\nu}{2}\,,
\end{align}
the variation of the local entropy $\ts$ is expressed as
\begin{align}
 \delta \ts = -\frac{u^\alpha}{T}\,\delta\tp_\alpha
  - \frac{\mu}{T}\,\delta\tn
  + \frac{\sqrt{h}}{2T}\,T_{\rm (q)}^{\mu\nu}\,\delta g_{\mu\nu}\,,
\label{delta_s}
\end{align}
where $T_{\rm (q)}^{\mu\nu}$ is the quasiconservative energy-momentum tensor:
\begin{align}
 T_{\rm (q)}^{\mu\nu} \equiv e\,u^\mu u^\nu + \tau_{\rm (q)}^{\mu\nu}
  = p^\mu u^\nu + \tau_{\rm (q)}^{\mu\nu}\,.
\end{align}
In order to make this decomposition unique, 
we require that $\tau_{\rm (q)}^{\mu\nu}$ be orthogonal to $u^\mu$\,:
\begin{align}
 \tau_{\rm (q)}^{\mu\nu}\,u_\nu = 0\,.
\end{align}
Simple fluids (that have no specific spatial directions) 
are realized by taking $\tau_{\rm (q)}^{\mu\nu}=P\,h^{\mu\nu}$ 
with $P$ the pressure. 
Then, by using the identity 
$h^{\mu\nu}\,\delta g_{\mu\nu}= h^{\mu\nu}\,\delta h_{\mu\nu} = 2\delta\sqrt{h}/\sqrt{h}$\,, 
Eq.\ \eq{delta_s} becomes the standard expression 
for the local entropy $\ts=\tsigma(\te,\tn,g_{\mu\nu})$ 
of simple fluids, 
$\delta \ts = \delta\tsigma =(1/T)\,\delta\te-(\mu/T)\,\delta\tn + P\,\delta\sqrt{h}$\,; 
or equivalently, $\delta s = \delta(\ts/\sqrt{h}) 
= (1/T)\,\delta e - (\mu/T)\,\delta n$\,.

As for the entropy functional, 
from the general expression \eq{total_entropy} without $b^A$-type variables, 
we have
\begin{align}
 \Delta\Stot &\equiv - \frac{1}{2} \int_{\Sigma_x[L_{\rm s}]} 
  \rmd^D\by\, N\sqrt{h}\,\Bigl[
  \ell^\mu_{~\alpha,}{}^\nu_{~\beta}\,
  \nabla_\mu\Bigl( \frac{\partial\ts}{\partial\tp_\alpha}\Bigr)\,
  \nabla_\nu\Bigl( \frac{\partial\ts}{\partial\tp_\beta}\Bigr)
  + m^{\mu\nu}\,\partial_\mu \Bigl(\frac{\partial\ts}{\partial\tn}\Bigr)\,
  \partial_\nu \Bigl(\frac{\partial\ts}{\partial\tn}\Bigr)\Bigr]\,.
\label{entropy_functional_fluid}
\end{align}
We assume that the coefficient functions $\ell^\mu_{~\alpha,}{}^\nu_{~\beta}$ 
and $m^{\mu\nu}$ are all orthogonal to the velocity field; 
$\ell^\mu_{~\alpha,}{}^\nu_{~\beta}\,u_\nu = 0$\,,
$\ell^\mu_{~\alpha,}{}^\nu_{~\beta}\,u^\beta=0$\,, and $m^{\mu\nu}\,u_\nu=0$\,,
and also that the second derivatives of the form 
$\partial^2\ts/\partial\tp_\alpha\partial\tn$ are small 
and can be neglected.  
The maximum of the entropy functional is then given by 
$(p^0_\alpha,n^0)$ that satisfy the following equations:%
\begin{align}
 h^{\mu\nu}\,h_{\alpha\beta}\,
  \nabla_\nu\Bigl(\frac{\partial \ts}{\partial \tp_\beta}\Bigr)\Bigr|_0
  &=h^{\mu\nu}\,h_{\alpha\beta}\,
  \nabla_\nu\Bigl(-\,\frac{u^\beta}{T}\Bigr)\Bigr|_0
  =-\frac{1}{T}\,h^{\mu\nu}\,h_{\alpha\beta}\,\nabla_\nu u^\beta\bigr|_0=0\,,\\
 h^{\mu\nu}\,\partial_\nu\Bigl(\frac{\partial \ts}{\partial \tn}\Bigr)\Bigr|_0
  &=h^{\mu\nu}\,\partial_\nu\Bigl(-\,\frac{\mu}{T}\Bigr)\Bigr|_0=0\,.
\end{align}
Note that these equations hold only 
within the spatial region $\Sigma_x[L_{\rm s}]$\,.

The variation of the entropy functional \eq{entropy_functional_fluid} 
is given by
\begin{align}
 \frac{\delta\Delta\Stot}{\delta\tp_\lambda} 
  &= \sqrt{-g}\,\frac{\partial^2\ts}{\partial\tp_\lambda\partial\tp_\alpha}\,
  \nabla_\mu\Bigl[ \ell^\mu_{~\alpha,}{}^\nu_{~\beta}\,\nabla_\nu\Bigl(
  \frac{\partial\ts}{\partial\tp_\beta}\Bigr)\Bigr]
  =\sqrt{-g}\,\frac{\partial^2\ts}{\partial\tp_\lambda\partial\tp_\alpha}\,
  \nabla_\mu\Bigl[ -\frac{1}{T}\,\ell^\mu_{~\alpha,}{}^\nu_{~\beta}\,\nabla_\nu u^\beta \Bigr]\,,\\
 \frac{\delta\Delta\Stot}{\delta\tn} 
  &= \sqrt{-g}\,\frac{\partial^2\ts}{\partial\tn^2}\,
  \nabla_\mu\Bigl[ m^{\mu\nu}\,\partial_\nu\Bigl(
  \frac{\partial\ts}{\partial\tn}\Bigr)\Bigr]
  = \sqrt{-g}\,\frac{\partial^2\ts}{\partial\tn^2}\,
  \nabla_\mu\Bigl[ m^{\mu\nu}\,\partial_\nu\Bigl(
  -\,\frac{\mu}{T}\Bigr)\Bigr]\,.
\label{s_nn}
\end{align}
By using the decomposition of the matrix 
$\partial^2\ts/\partial\tp_\mu\partial\tp_\nu$ 
(negative-definite for each irreducible component) as%
\begin{align}
 \sqrt{h}\,\frac{\partial^2 \ts}{\partial\tp_\mu \partial\tp_\nu}
  = -c_\bot&\,u^\mu u^\nu - c_\|\,h^{\mu\nu}\,, \nn\\
 \biggl(c_\bot = -\,\sqrt{h}\,\frac{\partial^2\ts}{\partial\te^2}\,(>0)\,,& \quad 
  c_\| = \frac{1}{e\, T} \,(>0)\biggr)\,,
\label{s_pp-decomp}
\end{align}
the vector $\delta\Delta\Stot/\delta \tp_\lambda$ 
is decomposed as
\begin{align}
 \frac{\delta \Delta\Stot}{\delta \tp_\lambda}
  &= \biggl[\frac{\delta \Delta\Stot}{\delta \tp_\lambda}
  \biggr]_{\mbox{\raisebox{3pt}{\scriptsize{$\bot$}}}}
  + \biggl[\frac{\delta \Delta\Stot}{\delta \tp_\lambda}
  \biggr]_{\mbox{\raisebox{3pt}{\scriptsize{$\|$}}}}
\end{align}
with
\begin{align}
 \biggl[\frac{\delta \Delta\Stot}{\delta \tp_\lambda}
  \biggr]_{\mbox{\raisebox{3pt}{\scriptsize{$\bot$}}}}
  &\equiv (-u^\lambda u^\nu)\,\frac{\delta \Delta\Stot}{\delta \tp_\nu}
  = N\,c_\bot\,(-u^\lambda u^\nu)\,
  \nabla_\mu\,\Bigl[-\, \frac{1}{T}\,\ell^\mu_{~\nu,}{}^\rho_{~\sigma}\,
  \nabla_\rho u^\sigma\Bigr] \,,
\label{s_pp_perp}\\
 \biggl[\frac{\delta \Delta\Stot}{\delta \tp_\lambda}
  \biggr]_{\mbox{\raisebox{3pt}{\scriptsize{$\|$}}}}
  &\equiv h^{\lambda\nu}\,\frac{\delta \Delta\Stot}{\delta \tp_\nu}
  = -\,N\,c_\|\,h^{\lambda\nu}\,
  \nabla_\mu\,\Bigl[-\, \frac{1}{T}\,\ell^\mu_{~\nu,}{}^\rho_{~\sigma}\,
  \nabla_\rho u^\sigma\Bigr] 
\label{s_pp_para}\,.
\end{align}
With the variations of the entropy functional 
Eqs.\ \eq{s_nn}, \eq{s_pp_perp}, and \eq{s_pp_para}, 
we set the following equations of linear regression:
\begin{align}
   \bigl[\,\dot{\tp}_\alpha -[\dot{\tp}_\alpha]_{\rm rev}\bigr]_{\bot}
 &=
 [\,\dot{\tp}_\alpha]_{{\rm irr}\,\bot}
  = \sqrt{h}\,L_\bot\,u_\alpha u_\lambda\,
  \biggl[\frac{\delta\Delta\Stot}{\delta\tp_\lambda}
  \biggr]_{\mbox{\raisebox{3pt}{\scriptsize{$\bot$}}}}\,,
\label{linear_regression1}\\
   \bigl[\,\dot{\tp}_\alpha -[\dot{\tp}_\alpha]_{\rm rev}\bigr]_{\|}
 &=
 [\,\dot{\tp}_\alpha]_{{\rm irr}\,\|}
   = \sqrt{h}\,L_\|\,h_{\alpha\lambda}\,
  \biggl[\frac{\delta \Delta\Stot}{\delta\tp_\lambda}
  \biggr]_{\mbox{\raisebox{3pt}{\scriptsize{$\|$}}}}\,,
\label{linear_regression2}\\
 \dot{\tn} - [\dot{\tn}]_{\rm rev}
 &=
 [\dot{\tn}]_{\rm irr}
  \equiv \sqrt{h}\,M\,\frac{\delta\Delta\Stot}{\delta\tn}\,.
\label{linear_regression3}
\end{align}

Note that the time derivatives of $\tp_\alpha$ and $\tn$ are given by%
\begin{align}
 \frac{\rmD}{\rmD t}\,\tp_\alpha &\equiv N\Lie_u\,\tp_\alpha
  = \sqrt{-g}\,\nabla_\mu\bigl( p_\alpha u^\mu\bigr)\,,
\label{def_D/Dt1}\\
 \frac{\rmD}{\rmD t}\,\tn &\equiv N\Lie_u\,\tn
  = \sqrt{-g}\,\nabla_\mu\bigl( n u^\mu)\,.
\label{def_D/Dt2}
\end{align}
Equations \eq{def_D/Dt1} and \eq{def_D/Dt2} can be shown by using 
the identities $\Lie_u\sqrt{h}=\sqrt{h}\,\nabla_\mu u^\mu$ 
and $p_\mu \nabla_\alpha u^\mu=0$\,. 
The former identity can be proved as  
$\Lie_u\sqrt{h}=(1/2)\sqrt{h}\,h^{\mu\nu}\Lie_u h_{\mu\nu}
=(1/2)$
$\sqrt{h}\,h^{\mu\nu}\Lie_u g_{\mu\nu}
=(1/2)\sqrt{h}\,h^{\mu\nu}\,(\nabla_\mu u_\nu+\nabla_\nu u_\mu)
=(1/2)\sqrt{h}\,g^{\mu\nu}\,(\nabla_\mu u_\nu+\nabla_\nu u_\mu)
=\sqrt{h}\,\nabla_\mu u^\mu$\,.
The latter follows from the facts that $p_\mu=e\,u_\mu$ and $u^\mu u_\mu=-1$\,.
One can easily show that 
the left-hand side of Eqs.\ \eq{linear_regression1}--\eq{linear_regression3} 
can be rewritten as
\begin{align}
   \bigl[\,\dot{\tp}_\alpha -[\dot{\tp}_\alpha]_{\rm rev}\,\bigr]_{\bot}
  &=(-u_\alpha\,u^\nu)\bigl[\,\sqrt{-g}\,\nabla_\mu(p_\nu\,u^\mu)
  - [\dot{\tp}_\nu]_{\rm rev}\,\bigr]\,,\\
   \bigl[\,\dot{\tp}_\alpha -[\dot{\tp}_\alpha]_{\rm rev}\,\bigr]_{\|}
  &= h_\alpha^{~\nu}\,\bigl[\,\sqrt{-g}\,\nabla_\mu(p_\nu\,u^\mu)
  - [\dot{\tp}_\nu]_{\rm rev}\,\bigr] \,,\\
 \dot{\tn} - [\dot{\tn}]_{\rm rev}
  &=\sqrt{-g}\,\nabla_\mu(n\,u^\mu) - [\dot{\tn}]_{\rm rev}\,.
\end{align}
Furthermore, one expects that the reversible (or isentropic) evolutions of $\tp_\mu$ 
and $\tn$ 
can be identified as the evolutions for ideal fluids, 
\begin{align}
 [\,\dot{\tp}_\alpha]_{\rm rev}
  = -\sqrt{-g}\,\nabla_\mu \tau^{\rm (r)}_\alpha{}^\mu\,,
  \quad [\dot{\tn}]_{\rm rev} = 0\,.
\label{ideal_fluid}
\end{align}
The isentropic current $\tau^{\rm (r)}_\alpha{}^\mu$
equals the quasiconservative stress tensor $\tau^{\rm (q)}_\alpha{}^\mu$, 
as is shown in the next section. 
Then Eqs.\ \eq{linear_regression1}--\eq{linear_regression3} become 
\begin{align}
 (-u_\alpha\,u^\nu)\,\nabla_\mu T^{\rm (q)}_\nu{}^\mu
  &=-\,(-u_\alpha\,u^\nu)\,c_\bot L_\bot\,\nabla_\mu\Bigl[
  -\frac{1}{T}\,\ell^\mu_{~\nu,}{}^\rho_{~\sigma}\,
  \nabla_\rho u^\sigma\Bigr]\,,
\label{current-1}\\
 h_\alpha^{~\nu}\,\nabla_\mu T^{\rm (q)}_\nu{}^\mu
  &=-\,h_\alpha^{~\nu}\,c_\| L_\|\,\nabla_\mu\Bigl[
  -\frac{1}{T}\,\ell^\mu_{~\nu,}{}^\rho_{~\sigma}\,
  \nabla_\rho u^\sigma\Bigr]\,,
\label{current-2}\\
 \nabla_\mu (n u^\mu)
  &= \sqrt{h}\,\frac{\partial^2\ts}{\partial\tn^2} \,M\,
  \nabla_\mu\Bigl[ m^{\mu\nu}\,\partial_\nu\Bigl(
  -\frac{\mu}{T}\Bigr)\Bigr]\,.
\label{current-3}
\end{align}
We see that general covariance requires that
$c_\bot\,L_\bot = c_\|\,L_\| \,(\equiv\,L_\tp)$\,. 
By assuming that $L_\tp$ 
and $L_\tn\equiv \sqrt{h}\,(-\partial^2\ts/\partial\tn^2)\,M$ 
can be regarded as being constant at this order of derivative expansion,
the linear regression is expressed as current conservations:
\begin{align}
 \nabla_\mu\,T_\nu^{~\mu}=0\,,\quad \nabla_\mu n^\mu=0\,,
\label{conservation1}
\end{align}
where the energy-momentum tensor and the particle-number current are given by
\begin{align}
 T_\nu^{~\mu} \equiv T^{\rm (q)}_\nu{}^{\mu} + \tau^{\rm (d)}_\nu{}^{\mu}
  =e\,u_\nu u^\mu +\tau^{\rm (q)}_\nu{}^{\mu} + \tau^{\rm (d)}_\nu{}^{\mu}\,,\quad
  n^\mu \equiv n\,u^\mu + \nu^\mu
\label{conservation2}
\end{align}
with
\begin{align}
 \tau^{\rm (d)}_\nu{}^{\mu} 
  \equiv -\frac{1}{T}\,L_\tp\,\ell^\mu_{~\nu,}{}^\rho_{~\sigma}
  \nabla_\rho u^\sigma\,,\quad
  \nu^\mu 
  \equiv L_\tn\,m^{\mu\nu}\,\partial_\nu\Bigl(
  -\frac{\mu}{T}\Bigr)\,.
\label{conservation3}
\end{align}

For simple fluids that have no specific directions, 
the quantities given above can be parametrized as  
\begin{align}
 \tau_{\rm (q)}^{\mu\nu}&=P\,h^{\mu\nu}\quad\mbox{($P$: pressure) }\,,\\
 \frac{1}{T}\,L_\tp\,\ell^{\mu\nu,\rho\sigma}
  &=\zeta\,h^{\mu\nu}\,h^{\rho\sigma}
   +\eta\,\bigl[h^{\mu\rho}\,h^{\nu\sigma}+h^{\mu\sigma}\,h^{\nu\rho}
    - (2/D)\, h^{\mu\nu}\,h^{\rho\sigma}\bigr]\,,\\
 L_\tn\,m^{\mu\nu}&= \sigma\,T^2\,h^{\mu\nu} \,,
\end{align}
where $\zeta\,,\ \eta$\,, and $\sigma$ $(\geq 0)$ 
are, respectively, the bulk viscosity, the shear viscosity, 
and the diffusion constant. 
Note that the contributions from the rotation 
$h_\mu^{~\rho}\,h_\nu^{~\sigma}\,\nabla_{[\rho}u_{\sigma]}$ 
have been discarded since it vanishes in the assumption 
that the velocity field $u^\mu$ is hypersurface orthogonal. 
Then Eqs.\ \eq{conservation1}--\eq{conservation3} become
the well-known conservation laws 
for viscous simple fluids in the Landau-Lifshitz frame:%
\begin{align}
 \nabla_\mu T^{\mu\nu}&=0\,,\qquad
 \nabla_\mu n^\mu =\nabla_\mu (n\,u^\mu + \nu^\mu) = 0 \,, 
\label{conservation_fluid}\\
  T^{\mu\nu} &\equiv e\,u^\mu u^\nu 
                  + \bigl(P-\zeta \nabla_\rho u^\rho\bigr)\,h^{\mu\nu} 
  - 2\eta \nabla^{\langle\mu} u^{\nu\rangle}\,,
\label{conservation_fluid2}\\
 \nu^\mu &\equiv 
 \sigma\,T^2\,h^{\mu\nu}\, \partial_\nu \Bigl(-\frac{\mu}{T}\Bigr) \,.
\label{conservation_fluid3}
\end{align}
Here and hereafter, we define 
$A^{\langle \mu\nu \rangle}\equiv 
(1/2)\,h^\mu_{~\rho}\,h^\nu_{~\sigma}\,\bigl( A^{\rho\sigma}+A^{\sigma\rho}
-(2/D)\, h_{\alpha\beta}\,A^{\alpha\beta}\,h^{\rho\sigma}\bigr)$ 
for a tensor $A^{\mu\nu}$\,.

\subsection{More on the entropy production}
\label{more_on_entropy_production}

The entropy production rate can also be calculated in the following way
once we know that the current conservations 
\eq{conservation_fluid}--\eq{conservation_fluid3}  hold.
First we consider the increase of the entropy functional 
during the time interval $\Delta t\,\,\,(\gtrsim \epsilon_{\rm s})$\,:
\begin{align}
 &\Stot(t+\Delta t;\,\Sigma_x[L_{\rm s}]) 
 -\Stot(t;\,\Sigma_x[L_{\rm s}]) 
  =\int_t^{t+\Delta t}\!\! \rmd t\,\dot{\Stot}(t;\,\Sigma_x[L_{\rm s}])\nn\\
 &=\int_t^{t+\Delta t}\!\! \rmd t \int_{\Sigma_x[L_{\rm s}]}\rmd^D\by\,
      \dot{\ts}\bigl(\tp_\mu(t,\by),\,\tn(t,\by),\,g_{\mu\nu}(t,\by)\bigr) \nn\\
 &=\int_t^{t+\Delta t}\!\!  \rmd t\int_{\Sigma_x[L_{\rm s}]}\rmd^D\by\,
   \Bigl[\,\frac{\partial \ts}{\partial \tp_\mu}\, \dot{\tp}_\mu
 +\frac{\partial \ts}{\partial \tn}\, \dot{\tn} 
 + \frac{\partial \ts}{\partial g_{\mu\nu}}\, \dot{g}_{\mu\nu}\,\Bigr] \nn\\
 &=\int_t^{t+\Delta t}\!\!\int_{\Sigma_x[L_{\rm s}]}\rmd^{D+1}y\,\sqrt{-g}\,\Bigl[\, 
 -\frac{u^\mu}{T} \, \nabla_\nu \bigl(p_\mu u^{\nu}\bigr)
 +\Bigl(-\frac{\mu}{T}\Bigr)\, \nabla_\mu\bigl(n u^{\mu}\bigr)
 +\frac{1}{T}\,T_{\rm (q)}^{\mu\nu}\, \nabla_\nu u_\mu \,\Bigr] \,.
\end{align}
Since $\nabla_\nu T^{\mu\nu}=0$ for $T^{\mu\nu}=p^\mu u^\nu+\tau^{\mu\nu}
\equiv p^\mu u^\nu+\tau_{\rm (q)}^{\mu\nu}+\tau_{\rm (d)}^{\mu\nu}$\,, 
$\nabla_\mu n^\mu=0$ for $n^\mu=n u^\mu+\nu^\mu$\,, 
and $T_{\rm (q)}^{\mu\nu}\,\nabla_\nu u_\mu
=\tau_{\rm (q)}^{\mu\nu}\,\nabla_\nu u_\mu
=-u_\mu\,\nabla_\nu\tau_{\rm (q)}^{\mu\nu}$\,, 
the above equation can be rewritten as
\begin{align}
 \Stot&(t+\Delta t;\,\Sigma_x[L_{\rm s}]) 
 -\Stot(t;\,\Sigma_x[L_{\rm s}]) \nn\\
 &=\int_t^{t+\Delta t}\!\!\int_{\Sigma_x[L_{\rm s}]}\rmd^{D+1}y\,\sqrt{-g}\,\Bigl[\, 
 \frac{u^\mu}{T} \, \nabla_\nu\bigl( \tau_\mu^{~\nu}-\tau_\mu^{{\rm (q)}\,\nu})
 +\frac{\mu}{T}\, \nabla_\mu \nu^{\mu} \,\Bigr] \nn\\
 &=\int_t^{t+\Delta t}\!\!\int_{\Sigma_x[L_{\rm s}]}\rmd^{D+1}y\,\sqrt{-g}\,\Bigl[\, 
   \frac{u^\mu}{T} \,\nabla_\nu \tau_\mu^{{\rm (d)}\,\nu}
 + \frac{\mu}{T} \, \nabla_\mu \nu^{\mu} \,\Bigr] \,.
\label{entropy_dot}
\end{align}
The first equality means that 
$\tau_\mu^{~\nu}$ equals $\tau_\mu^{{\rm (q)}\,\nu}$ 
in the absence of dissipation, 
and thus shows that the equality 
$\tau_\mu^{{\rm (r)}\,\nu}=\tau_\mu^{{\rm (q)}\,\nu}$ 
holds for fluids.
By integrating by parts, the equation can be further rewritten as
\begin{align}
 \Stot&(t+\Delta t;\,\Sigma_x[L_{\rm s}]) 
 -\Stot(t;\,\Sigma_x[L_{\rm s}]) \nn\\
 &=\int_t^{t+\Delta t}\!\!\int_{\Sigma_x[L_{\rm s}]}\rmd^{D+1}y\,\sqrt{-g}\,\Bigl[\, 
  - \frac{1}{T} \, \tau_{\rm (d)}^{\mu\nu}\,K_{\mu\nu}
 + \nu^{\mu}\,\partial_\mu \Bigl(-\frac{\mu}{T}\Bigr)
 + \nabla_\mu \Bigl(\frac{\mu}{T}\, \nu^{\mu}\Bigr)\,\Bigr] \,.
\end{align}
The last term on the right-hand side is the effect from the surroundings 
so that the entropy production in the bulk is given by
\begin{align}
 \Stot&(t+\Delta t;\,\Sigma_x[L_{\rm s}]) 
 -\Stot(t;\,\Sigma_x[L_{\rm s}]) 
    -\int_t^{t+\Delta t} \int_{\Sigma_x[L_{\rm s}]}\rmd^{D+1}y\,\sqrt{-g}\,
   \nabla_\mu \Bigl(\frac{\mu}{T}\, \nu^{\mu}\Bigr)\nn\\
 &=\int_t^{t+\Delta t}\!\! \int_{\Sigma_x[L_{\rm s}]}\rmd^{D+1}y\,\sqrt{-g} \,\Bigl[\, 
  - \frac{1}{T} \, \tau_{\rm (d)}^{\mu\nu}\,K_{\mu\nu}
 + \nu^{\mu}\,\partial_\mu \Bigl(-\frac{\mu}{T}\Bigr) \,\Bigr] \,.
\label{entropy_production_fluid}
\end{align}
Since $\dot\ts$ can be expressed as $\sqrt{-g}\,\nabla_\mu( s\,u^\mu)$\,,
the left-hand side of Eq.\ \eq{entropy_production_fluid} can be written as
$\int_t^{t+\Delta t}\int_{\Sigma_x[L_{\rm s}]}
\rmd^{D+1}y\,\sqrt{-g}\,\nabla_\mu\bigl[s\,u^\mu-(\mu/T)\,\nu^\mu\bigr]$\,.
Thus, by defining the entropy current as 
$s^\mu \equiv s\,u^\mu-(\mu/T)\, \nu^{\mu}$\,, 
Eq.\ \eq{entropy_production_fluid} can be expressed in a local form as
\begin{align}
 \nabla_\mu s^\mu &=
  - \frac{1}{T} \, \tau_{\rm (d)}^{\mu\nu}\,K_{\mu\nu}
  +\nu^{\mu}\,\partial_\mu \Bigl(-\frac{\mu}{T}\Bigr) \nn\\
  &= \frac{\zeta}{T}\,(h^{\mu\nu} K_{\mu\nu})^2 + \frac{2\eta}{T}\,
       K^{\langle\mu\nu\rangle}\,K_{\langle\mu\nu\rangle}
  + \sigma\,T^2\,
        h^{\mu\nu}\,\partial_\mu \Bigl(-\frac{\mu}{T}\Bigr)\,
                    \partial_\nu \Bigl(-\frac{\mu}{T}\Bigr) \,.
\label{entropy_divergence}
\end{align}

Equation \eq{entropy_divergence} might seem to be inconsistent with the 
entropy production rate given in Eq.\ \eq{entropy_production},
because Eq.\ \eq{entropy_production} predicts that 
the entropy production rate is the spatial integral of
a quadratic form in $\nabla_\mu \nu^\mu$ 
and $\nabla_\mu \tau_{\rm (d)}^{\mu\nu}$\,.
However, they are actually consistent.
In fact, suppose that we expand $u^\mu/T$ and $\mu/T$ in Eq.\ \eq{entropy_dot} 
around their equilibrium values
$(u^\mu_0/T_0)(y;\,L_{\rm s})$ and $(\mu_0/T_0)(y;\,L_{\rm s})$ as
\begin{align}
 \frac{u^\mu}{T}=\frac{u_0^\mu}{T_0}
  +\Bigl(\frac{u^\mu}{T} - \frac{u_0^\mu}{T_0}\Bigr)\,,\quad
  \frac{\mu}{T}=\frac{\mu_0}{T_0}
  +\Bigl(\frac{\mu}{T} - \frac{\mu_0}{T_0}\Bigr)\,.
\end{align}
Since the deviations $u^\mu/T-u_0^\mu/T_0$ and $\mu/T-\mu_0/T_0$ 
are proportional to their thermodynamic forces, 
$\delta\Delta\Stot/\delta \tp_\mu$ 
and $\delta\Delta\Stot/\delta \tn$\,, respectively, 
they can be expressed in the following form:
\begin{align}
 \frac{u^\mu}{T} - \frac{u_0^\mu}{T_0} 
  =- \nabla_\nu \Bigl[\gamma^{\mu\nu,\,\rho\sigma}\,
                            \nabla_\rho \Bigl(\frac{u_\sigma}{T}\Bigr)\Bigr]\,,\quad
  \frac{\mu}{T} - \frac{\mu_0}{T_0} 
  =- \nabla_\mu \Bigl[\, \gamma^{\mu\nu}\, \partial_\nu \Bigl(\frac{\mu}{T}\Bigr)\,\Bigr]\,,
\end{align}
where $\gamma^{\mu\nu,\,\rho\sigma}$ and $\gamma^{\mu\nu}$ 
are some positive semidefinite matrices. 
Substituting them into Eq.\ \eq{entropy_dot}, 
we have the following equation:
\begin{align}
 &\dot{\Stot}(t;\,\Sigma_x[L_{\rm s}])
 -\int_{\Sigma_x[L_{\rm s}]}\!\!\rmd^D\by\, \sqrt{-g}\,
  \nabla_\nu \Bigl(\,\frac{u_0^\mu}{T_0}\,\tau_\mu^{{\rm (d)}\,\nu}
 + \frac{\mu_0}{T_0}  \, \nu^{\nu}\,\Bigr) \nn\\
 &=\int_{\Sigma_x[L_{\rm s}]}\!\!\rmd^D\by\,\sqrt{-g}\, 
 \Bigl[- \nabla_\nu 
        \bigl[\bigl(\gamma^{\mu\nu,\,\rho\sigma}/T\bigr) K_{\rho\sigma}\bigr]\, 
       \nabla_\lambda \tau_\mu^{{\rm (d)}\,\lambda}
 - \nabla_\mu \bigl[\, \gamma^{\mu\nu} \partial_\nu \bigl(\mu/T \bigr)\bigr] \, 
   \nabla_\rho \nu^{\rho} \, \Bigr] \,.
\label{entropy_prod}
\end{align}
Here $(u_0^\mu/T_0)\,\tau_\mu^{{\rm (d)}\,\nu}$ can be neglected 
at this order of approximation 
because $u_0^\mu\,\tau_\mu^{{\rm (d)}\,\nu}\simeq 
u^\mu\,\tau_\mu^{{\rm (d)}\,\nu}=0$\,, 
while $\nabla_\nu [(\mu_0/T_0)\, \nu^{\nu}]$ 
can be replaced by  $\nabla_\nu [(\mu/T)\, \nu^{\nu}]$\,.
Thus, the second term on the left-hand side of Eq.\ \eq{entropy_prod} 
can be written as $-\int_{\Sigma_x[L_{\rm s}]}\rmd^D\by\,\sqrt{-g}\, 
\nabla_\nu \bigl[(\mu/T)\,\nu^\nu\bigr]$\,, 
so that the left-hand side of Eq.\ \eq{entropy_prod}
agrees with that of Eq.\ \eq{entropy_production_fluid}.
We thus find that the entropy production rate in the bulk
can also be written as a quadratic form in $\nabla_\mu \nu^\mu$ 
and $\nabla_\mu \tau_{\rm (d)}^{\mu\nu}$\,.

\section{Relativistic viscoelastic fluid mechanics}
\label{rel_visc_mech}

\subsection{Intrinsic metric}
\label{intrinsic_metric}

Once the motion of a material is specified by a velocity field $u^\mu(x)$\,, 
the shape of the timeslice normal to $u^\mu$ 
is represented by the induced metric \eq{hypersurface_metric}, 
$h_{\mu\nu} = g_{\mu\nu}+u_\mu u_\nu$\,.
If the relaxation time for the deformation of shape can be regarded 
as being very short, 
then, as discussed in Sec.\ \ref{relativistic_fluid_mecanics},
the fundamental relation, $\ts(x)=\tsigma\bigl(\te(x),\tn(x),g_{\mu\nu}(x)\bigr)$\,,
does not depend on the hysteresis of the shape $h_{\mu\nu}(x)$\,.
Almost all materials (besides fluids), however, 
do not have such short relaxation times for the deformation of shape, 
and the time evolution of shape may not be described as a Markovian process; 
all the preceding history needs to be known 
in order to predict the future behavior of the system at a given initial timeslice. 
One then needs to extend the formalism 
such that the equations contain time derivatives of higher orders. 
This is out of the scope of the standard nonequilibrium thermodynamics.

However, a class of materials still allow a standard thermodynamic description 
even when their relaxation time of shape is not short. 
This is performed by replacing the introduction of higher time derivatives 
with that of extra dynamical degrees of freedom. 
Viscoelastic materials considered below 
belong to this class of materials.

\begin{figure}[htbp]
\vspace{3ex}
\begin{center}
\includegraphics[scale=0.6]{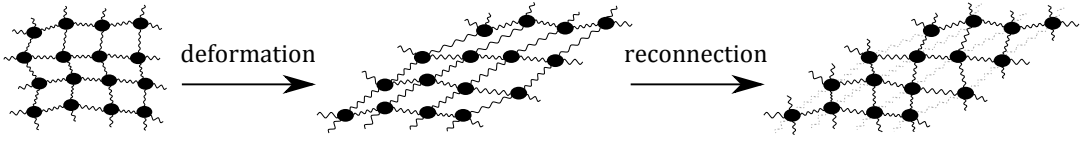}
\begin{quote}
\caption{Processes of deformation and strain relaxation.\label{fig:reconnection}}
\end{quote}
\end{center}
\vspace{-6ex}
\end{figure}

According to the definition of Maxwell, 
viscoelastic materials behave as elastic solids at short time scales 
and as viscous fluids at long time scales 
(see, e.g., Sec.\ 36 in \cite{LL_elasticity}). 
In order to understand how such materials evolve in time, 
we consider a material consisting of many molecules bonding each other 
and assume that the molecules first stay at their equilibrium positions 
in the absence of strains 
(as in the leftmost illustration of Fig.\ \ref{fig:reconnection}) \cite{afky,afky2}. 
We now suppose that an external force is applied to deform the material.
An internal strain is then produced in the body, and according to the definition, 
the accompanied internal stress can be treated as an elastic force, 
at least during short intervals of time.
However, if we keep the deformation much longer than the relaxation time
(characteristic to each material), 
then the bonding structure changes to maximize the entropy, 
and the internal strain vanishes eventually 
as in the rightmost of Fig.\ \ref{fig:reconnection}.
The point is that two figures (the central and the rightmost) 
have the same shape (same induced metric) $h_{\mu\nu}$\,, 
but different bonding structures.

\begin{figure}[htbp]
\vspace{3ex}
\begin{center}
\includegraphics[scale=0.55]{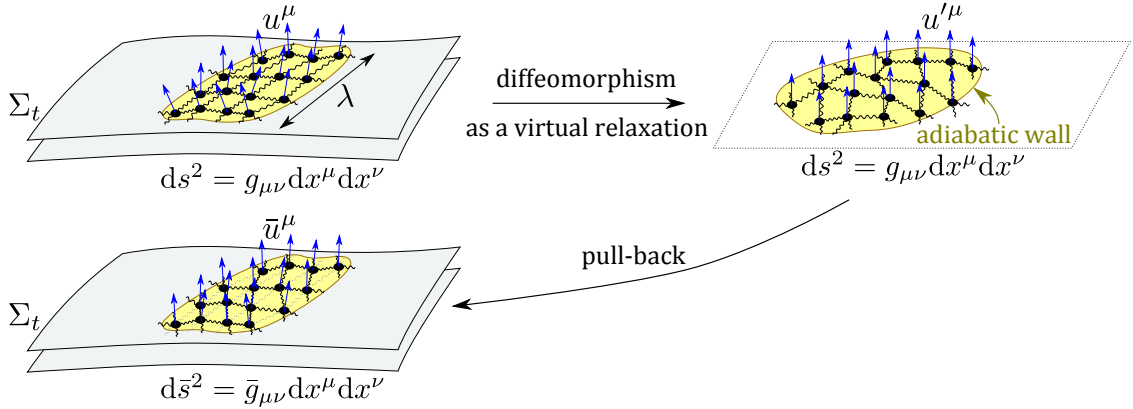}
\begin{quote}
\caption{The procedure to obtain the intrinsic metric.} \label{fig:intrinsic}
\end{quote}
\end{center}
\vspace{-6ex}
\end{figure}

In order to describe the internal bonding structure, 
we introduce at each moment 
another dynamical variable to be called the {\it intrinsic metric}, 
which measures the shape that the material would take 
when all the internal strains are removed virtually \cite{Eckart:1948,afky,afky2}. 
To define such states, 
around each spacetime point $x$\,, 
we consider a small spatial subregion 
whose linear size $\lambda$ is much smaller than $L_{\rm s}$ 
and enclose the subregion with an adiabatic and impermeable wall 
of vanishing tension (see Fig.\ \ref{fig:intrinsic}). 
We then cut it out of the bulk $\Sigma_x[L_{\rm s}]$ 
and leave it for a long time (comparable to $L_{\rm t}$). 
The subregion will then undergo a relaxation to reach its equilibrium state. 
This causes a diffeomorphism in $(D+1)$-dimensional spacetime 
from the subregion to another subregion, 
mapping the original (or real) positions of material particles 
to their new positions in the virtual equilibrium, 
$f_\lambda:\,x\mapsto f_\lambda(x)$\,. 
Since the absolute value of the extrinsic curvature gets reduced 
in the course of relaxation, 
this diffeomorphism does not necessarily preserve the foliation, 
and the velocity field $u_\mu(x)$ will change as  
$u_\mu(x)\to u_\mu'(x)$\,.%
We then consider the pullbacks of the resulting fields 
in the virtual equilibrium:
\begin{align}
 \bar{u}_\mu(x;\,\lambda) \equiv (f_\lambda^\ast u')_\mu(x)\,, \quad
  \bar{g}_{\mu\nu}(x;\,\lambda) \equiv (f_\lambda^\ast g)_{\mu\nu}(x)\,,
\end{align}
and define the intrinsic velocity field $\bar{u}_\mu(x)$ 
and the intrinsic metric $\bar{g}_{\mu\nu}(x)$ 
as the values in the limit $\lambda\to 0$\,:
\begin{align}
 \bar{u}_\mu(x) \equiv \lim_{\lambda\to 0}\,(f_\lambda^\ast u')_\mu(x)\,, \quad
  \bar{g}_{\mu\nu}(x) \equiv \lim_{\lambda\to 0}\,(f_\lambda^\ast g)_{\mu\nu}(x)\,.
\end{align}

Since $\bar{g}^{\mu\nu}\,\bar{u}_\nu$ is no longer 
orthogonal to the original hypersurface, 
the intrinsic metric $\barg_{\mu\nu}(x)$ may have tilted components.
We parametrize $\barg_{\mu\nu}$ as
\begin{align}
 \barg_{\mu\nu} &= -\,(1+2\theta)\,u_\mu u_\nu
  -\,\varepsilon_\mu u_\nu -\,\varepsilon_\nu\, u_\mu
  + \barh_{\mu\nu}\nn\\
 &\bigl( 
  \varepsilon_\mu u^\mu=0\,,\quad h_{\mu\nu}\, u^\nu=0\,,
  \quad \barh_{\mu\nu}\, u^\nu=0\bigr)\,.
\end{align}
The quantity $\sqrt{1+2\theta}$ represents the ratio of the temperature 
in the presence of strains 
to that in the absence of strains, 
because the Tolman law ($\sfT=N T=\bar{N}\bar{T}$)
tells that $T/\barT=\barN/N=\sqrt{1+2\theta}$\,, 
that is, 
$\theta=(T^2-\barT^2)/2\barT^2\simeq (T-\barT)/\barT$\,. 
The quantity $\varepsilon_{\mu}$ represents the relative velocity 
of the real path of a material particle to that of its virtual path. 
The spatial metric $\barh_{\mu\nu}$ is the same 
with the ``strain metric''
introduced by Eckart to embody 
``the principle of relaxability-in-the-small'' in anelasticity \cite{Eckart:1948}. 
(This was reinvented in \cite{afky} in the light of the covariant description 
of viscoelasticity  
under the foliation preserving diffeomorphisms.)
We further introduce the {\it strain tensor} 
\begin{align}
 E_{\mu\nu}(x) 
  &\equiv \frac{1}{2}\,\bigl(g_{\mu\nu}(x)-\barg_{\mu\nu}(x)\bigr) \nn\\
 &= \,\theta\, u_\mu u_\nu 
 + \frac{1}{2}\,\bigl(\varepsilon_\mu u_\nu
 + \varepsilon_\nu\, u_\mu \bigr)+ \varepsilon_{\mu\nu}\,,
\label{strain_general}
\end{align}
where 
\begin{align}
 \varepsilon_{\mu\nu}(x)
  \equiv \frac{1}{2}\,\bigl(h_{\mu\nu}(x)-\barh_{\mu\nu}(x) \bigr)
\label{spatial_strain}
\end{align}
is the spatial strain tensor.
The explicit forms of $h_{\mu\nu}$ and $\barh_{\mu\nu}$ 
under various deformations can be found in \cite{afky2}.
Note that if we define the extrinsic curvature
associated with the spatial intrinsic metric $\barh_{\mu\nu}$ as
\begin{align}
 \barK_{\mu\nu} \equiv \frac{1}{2}\,\Lie_u \barh_{\mu\nu}
  =\frac{1}{2}\,\bigl(u^\lambda\, \partial_\lambda \barh_{\mu\nu}
  + \partial_\mu u^\lambda\, \barh_{\lambda\mu}
  + \partial_\nu u^\lambda\, \barh_{\mu\lambda}\bigr)\,,
\end{align}
the following identity holds:
\begin{align}
 \Lie_u\varepsilon_{\mu\nu} = K_{\mu\nu}-\barK_{\mu\nu}\,.
\end{align}
We denote the contraction of a spatial tensor $A_{\mu\nu}$ 
with $g^{\mu\nu}$ by $\tr A$ so that
\begin{align}
 \tr\varepsilon\equiv g^{\mu\nu}\varepsilon_{\mu\nu}=h^{\mu\nu}\varepsilon_{\mu\nu}\,,
  \quad \tr K\equiv g^{\mu\nu}K_{\mu\nu}=h^{\mu\nu}K_{\mu\nu}\,.
\end{align}

\subsection{Linear regression and the rheology equations}
\label{linear_regression_rheology_eq}

We now apply the framework of section \ref{general_theory} 
with the following identifications:
\begin{center}
\begin{tabular}{|c|c||c|c||c|c|}
\hline
 $b^A$ & $b^A_0$ & 
 $\tc^I$  & $\beta_I$ & $d^P$ & $(\partial\ts/\partial d^P)$ \\ \hline\hline
 $\varepsilon_{\mu\nu}$& $0$ & 
 $\tp_\mu$ & $ - u^\mu/T$ & $g_{\mu\nu}$ & $\sqrt{h}\,T_{\rm (q)}^{\mu\nu}/2T$ \\
 $\varepsilon_\mu$& $0$ & 
 $\tn$ & $-\mu/T$ & & \\
 $\theta$& $0$ & & & & \\
\hline
\end{tabular}
\end{center}
Following the argument used in deriving Eq.\ \eq{entropy_functional_fluid},
the entropy functional \eq{total_entropy} are generically
expanded as follows:%
\begin{align}
 \,\Delta\Stot&(t;\,\Sigma_x[L_{\rm s}])\nn\\
 =&\,-\frac{1}{2}\int_{\Sigma_x[L_{\rm s}]} \!\!\!\rmd^D\by\, \sqrt{-g}\,\Biggl[
 \begin{pmatrix}
  \varepsilon_{\langle\mu\nu\rangle} & 
  \nabla^{\langle \mu} \bigl(\partial\ts/\partial \tp_{\nu\rangle}\bigr)
 \end{pmatrix} 
  \begin{pmatrix}
  \ell_1^{\langle\mu\nu\rangle,\langle\rho\sigma\rangle} 
 &\ell^{\langle\mu\nu\rangle,}_{2~~~\langle\rho\sigma\rangle} \cr
  \ell_{2\,\langle\mu\nu\rangle,}^{~~~~~\langle\rho\sigma\rangle}
 & \ell_{3\,\langle\mu\nu\rangle,\langle\rho\sigma\rangle}
  \end{pmatrix}
 \begin{pmatrix}
  \varepsilon_{\langle\rho\sigma\rangle} \cr 
  \nabla^{\langle \rho} \bigl(\partial\ts/\partial \tp_{\sigma\rangle}\bigr)
 \end{pmatrix} \nn\\
 &\qquad\qquad\qquad\quad\quad 
 +\begin{pmatrix}
  \varepsilon_{\mu} & 
  \partial_\mu \bigl(\partial\ts/\partial\tn\bigr)
 \end{pmatrix} 
  \begin{pmatrix}
  \ell_1^{\mu\nu} & \ell_2^{\mu\nu}\cr
  \ell_2^{\mu\nu} & \ell_3^{\mu\nu}
  \end{pmatrix}
 \begin{pmatrix}
  \varepsilon_{\nu} \cr 
  \partial_\nu \bigl(\partial\ts/\partial\tn\bigr)
 \end{pmatrix} \nn\\
 &\qquad\qquad\qquad\quad\quad 
 +\begin{pmatrix}
    \tr \varepsilon & \theta & \nabla_\mu\bigl(\partial\ts/\partial\tp_\mu\bigr)
 \end{pmatrix} 
  \begin{pmatrix}
  \ell^{\rm s}_1 & \ell^{\rm s}_2 & \ell^{\rm s}_4 \cr
  \ell^{\rm s}_2 & \ell^{\rm s}_3 & \ell^{\rm s}_5 \cr
  \ell^{\rm s}_4 & \ell^{\rm s}_5 & \ell^{\rm s}_6
  \end{pmatrix}
 \begin{pmatrix}
    \tr \varepsilon \cr \theta \cr \nabla_\mu\bigl(\partial\ts/\partial\tp_\mu\bigr)
 \end{pmatrix} \Biggr]  \nn\\
 =&\,-\frac{1}{2}\int_{\Sigma_x[L_{\rm s}]} \!\!\!\rmd^D\by\, \sqrt{-g}\,\Biggl[
 \begin{pmatrix}
  \varepsilon_{\langle\mu\nu\rangle} & 
  (-1/T)K_{\langle\mu\nu\rangle}
 \end{pmatrix} 
  \begin{pmatrix}
  \ell_1^{\langle\mu\nu\rangle,\langle\rho\sigma\rangle} 
 &\ell_2^{\langle\mu\nu\rangle,\langle\rho\sigma\rangle}\cr
  \ell_2^{\langle\mu\nu\rangle,\langle\rho\sigma\rangle} 
 & \ell_3^{\langle\mu\nu\rangle,\langle\rho\sigma\rangle}
  \end{pmatrix}
 \begin{pmatrix}
  \varepsilon_{\langle\rho\sigma\rangle} \cr 
  (-1/T)K_{\langle\rho\sigma\rangle}
 \end{pmatrix} \nn\\
 &\qquad\qquad\qquad\quad\quad 
 +\begin{pmatrix}
  \varepsilon_{\mu} & 
  \partial_\mu (-\mu/T)
 \end{pmatrix} 
  \begin{pmatrix}
  \ell_1^{\mu\nu} & \ell_2^{\mu\nu}\cr
  \ell_2^{\mu\nu} & \ell_3^{\mu\nu}
  \end{pmatrix}
 \begin{pmatrix}
  \varepsilon_{\nu} \cr 
  \partial_\nu (-\mu/T)
 \end{pmatrix} \nn\\
 &\qquad\qquad\qquad\quad\quad 
 +\begin{pmatrix}
    \tr \varepsilon & \theta & (-1/T)\tr K
 \end{pmatrix} 
  \begin{pmatrix}
  \ell^{\rm s}_1 & \ell^{\rm s}_2 & \ell^{\rm s}_4 \cr
  \ell^{\rm s}_2 & \ell^{\rm s}_3 & \ell^{\rm s}_5 \cr
  \ell^{\rm s}_4 & \ell^{\rm s}_5 & \ell^{\rm s}_6
  \end{pmatrix}
 \begin{pmatrix}
    \tr \varepsilon \cr \theta \cr (-1/T)\tr K
 \end{pmatrix} \Biggr]  \,,
\label{total_entropy2}
\end{align}
where the coefficient matrices are symmetric and positive semidefinite, 
and their indices are taken to be all orthogonal to $u^\mu$\,. 
Note that we again have used the fact that the rotation 
$h_\mu^{~\rho}\,h_\nu^{~\sigma}\,\nabla_{[\rho}u_{\sigma]}$ vanishes here. 
Since the matrices must be invariant tensors, 
we can assume that they take the following form:
\begin{align}
  \begin{pmatrix}
  \ell_1^{\langle\mu\nu\rangle,\langle\rho\sigma\rangle} 
 &\ell_2^{\langle\mu\nu\rangle,\langle\rho\sigma\rangle}\cr
  \ell_2^{\langle\mu\nu\rangle,\langle\rho\sigma\rangle} 
 & \ell_3^{\langle\mu\nu\rangle,\langle\rho\sigma\rangle}
  \end{pmatrix}
 &= 
  \begin{pmatrix}
  \ell_1^{\rm t} 
 &\ell_2^{\rm t} \cr
  \ell_2^{\rm t} 
 & \ell_3^{\rm t} 
  \end{pmatrix}
  \,h^{\langle\mu}_{\mu'}\, h^{\nu\rangle}_{\nu'}\, h^{\mu'\rho}\,h^{\nu'\sigma} \,, \\ 
  \begin{pmatrix}
  \ell_1^{\mu\nu} & \ell_2^{\mu\nu}\cr
  \ell_2^{\mu\nu} & \ell_3^{\mu\nu}
  \end{pmatrix}
 &=
    \begin{pmatrix}
  \ell_1^{\rm v} & \ell_2^{\rm v}\cr
  \ell_2^{\rm v} & \ell_3^{\rm v}
  \end{pmatrix} \, h^{\mu\nu}\,,
\end{align}
where 
$\Bigl(\begin{smallmatrix}
  \ell_1^{\rm t} 
 &\ell_2^{\rm t} \cr
  \ell_2^{\rm t} 
 & \ell_3^{\rm t} 
\end{smallmatrix}\Bigr)$ and 
$\Bigl(\begin{smallmatrix}
  \ell_1^{\rm v} & \ell_2^{\rm v}\cr
  \ell_2^{\rm v} & \ell_3^{\rm v}
\end{smallmatrix}\Bigr)$
are positive semidefinite.
Note that $\ell_k^{\langle\mu\nu\rangle,\langle\rho\sigma\rangle}\,
\varepsilon_{\langle\rho\sigma\rangle}
 = \ell^{\rm t}_k\,\varepsilon^{\langle\mu\nu\rangle}$ $(k=1,2,3)$\,. 
The functional derivatives are then evaluated to be
\begin{align}
 \frac{\delta \Delta\Stot}{\delta \varepsilon_{\langle\mu\nu\rangle}}
  &= -\, \sqrt{-g}\,\bigl[\,\ell_1^{\rm t}\,\varepsilon^{\langle\mu\nu\rangle} 
  -(\ell_2^{\rm t}/T)\,K^{\langle\mu\nu\rangle}\,\bigr] \,,
\label{varepsilon_munu}\\
 \frac{\delta \Delta\Stot}{\delta \varepsilon_{\mu}}
  &= -\, \sqrt{-g}\,h^{\mu\nu}\bigl[\,\ell_1^{\rm v}\, \varepsilon_{\nu}
             + \ell_2^{\rm v}\, \partial_\nu (-\mu/T)\,
       \bigr]
  ~~\biggl( =\biggl[ \frac{\delta \Delta\Stot}{\delta \varepsilon_{\mu}}
  \biggr]_{\mbox{\raisebox{3pt}{\scriptsize{$\|$}}}} \biggr)
 \,,
\label{varepsilon_mu}\\
 \frac{\delta \Delta\Stot}{\delta (\tr \varepsilon)}
  &=  -\, \sqrt{-g}\,\bigl[\,\ell_1^{\rm s}\, \tr \varepsilon
             + \ell_2^{\rm s}\, \theta
             - (\ell_4^{\rm s}/T)\, \tr K
       \,\bigr] \,,
\label{tr_varspsilon}\\
 \frac{\delta \Delta\Stot}{\delta \theta}
  &=  -\, \sqrt{-g}\,\bigl[\,\ell_2^{\rm s}\, \tr \varepsilon
             + \ell_3^{\rm s}\, \theta
             - (\ell_5^{\rm s}/T)\, \tr K
       \,\bigr] \,,
\label{theta}\\
 \frac{\delta \Delta\Stot}{\delta \tp_\mu}
  &= -\,N\,\bigl(c_\bot\,u^\mu u^\nu+c_\|\, h^{\mu\nu}\bigr)\times\nn\\
 &~~~~~ \times \nabla^\rho \bigl[\,
  \ell_2^{\rm t} \varepsilon_{\langle\nu\rho\rangle}
  -(\ell_3^{\rm t}/T) K_{\langle\nu\rho\rangle}
    + \bigl(\ell_4^{\rm s}\,\tr \varepsilon
    +\ell_5^{\rm s}\,\theta
    -(\ell_6^{\rm s}/T)\,\tr K\bigr) h_{\nu\rho} \,\bigr]\nn\\
 &=\biggl[ \frac{\delta \Delta\Stot}{\delta \tp_\mu}
  \biggr]_{\mbox{\raisebox{3pt}{\scriptsize{$\bot$}}}}
  +\biggl[ \frac{\delta \Delta\Stot}{\delta \tp_\mu}
  \biggr]_{\mbox{\raisebox{3pt}{\scriptsize{$\|$}}}}\,,
\label{pt_mu}\\
 \frac{\delta \Delta\Stot}{\delta \tn}
  &= -\,\sqrt{-g}\,\Bigl(-\frac{\partial^2\ts}{\partial\tn^2}\Bigr)
  \,\nabla_\mu \bigl[\,
 \ell_2^{\rm v}\, h^{\mu\nu}\,\varepsilon_{\nu}
  + \ell_3^{\rm v}\, h^{\mu\nu}\,\partial_\nu (-\mu/T)
  \bigr]\,,
\label{tn}
\end{align}
where we have made the following parametrization as in Eq.\ \eq{s_pp-decomp}:
\begin{align}
 \sqrt{h}\,\frac{\partial^2\ts}{\partial\tp_\mu\partial\tp_\nu}\Bigr\rvert_0
  = -\, c_\bot\,u^\mu u^\nu - c_\|\,h^{\mu\nu}\,.
\end{align}
Note that $\delta\Delta\Stot/\delta\varepsilon_\mu$ has 
only the components tangent to the timeslices.

The equations of linear regression are given 
in the following form:
\begin{align}
 \bigl[\dot\varepsilon_{\langle\mu\nu\rangle}\bigr]_{\rm irr}
  &\equiv \frac{1}{\sqrt{h}}\,
  L^{\varepsilon_{\langle\mu\nu\rangle}\varepsilon_{\langle\rho\sigma\rangle}}\,
  \frac{\delta\Delta\Stot}{\delta\varepsilon_{\rho\sigma}}\,,
\label{eq_varepsilon_munu}\\
 \bigl[\dot\varepsilon_{\mu}\bigr]_{{\rm irr}\,\bot}
  &\equiv \frac{1}{\sqrt{h}}\,
  L_\bot^{\varepsilon_\mu \varepsilon_\nu}\,
  \biggl[\frac{\delta\Delta\Stot}{\delta\varepsilon_\nu}
  \biggr]_{\mbox{\raisebox{3pt}{\scriptsize{$\bot$}}}} \equiv 0\,,
\label{eq_varepsilon_mu_perp}\\
 \bigl[\dot\varepsilon_{\mu}\bigr]_{{\rm irr}\,\|}
  &\equiv \frac{1}{\sqrt{h}}\,
  L_\|^{\varepsilon_\mu \varepsilon_\nu}\,
  \biggl[\frac{\delta\Delta\Stot}{\delta\varepsilon_\nu}
  \biggr]_{\mbox{\raisebox{3pt}{\scriptsize{$\|$}}}}\,,
\label{eq_varepsilon_mu_para}\\
 \begin{pmatrix}
  \bigl[(\tr\varepsilon)^\cdot\bigr]_{\rm irr} \cr
  \bigl[\,\dot\theta\,\bigr]_{\rm irr}
  \end{pmatrix}
  &\equiv \frac{1}{\sqrt{h}}\,
  \begin{pmatrix}
   L^{\tr\varepsilon\,\tr\varepsilon} & L^{\tr\varepsilon\,\theta}\cr
   L^{\tr\varepsilon\,\theta} & L^{\theta\,\theta}\
  \end{pmatrix}
  \begin{pmatrix}
   \delta\Delta\Stot/\delta (\tr\varepsilon) \cr
   \delta\Delta\Stot/\delta \theta
  \end{pmatrix}\,,
\label{eq_scalar}\\
 \bigl[\,\dot\tp_{\mu}-[\dot\tp_\mu]_{\rm rev}\bigr]_\bot
 &\equiv
 \bigl[\,\dot\tp_{\mu}\bigr]_{{\rm irr}\,\bot}
  \equiv 
  \sqrt{h}\,L_\bot^{\tp_\mu \tp_\nu}\,
  \biggl[\frac{\delta\Delta\Stot}{\delta\tp_\nu}
  \biggr]_{\mbox{\raisebox{3pt}{\scriptsize{$\bot$}}}}\,,
\label{eq_tp_mu_perp}\\
 \bigl[\,\dot\tp_{\mu}-[\dot\tp_\mu]_{\rm rev}\bigr]_\|
 &\equiv
 \bigl[\,\dot\tp_{\mu}\bigr]_{{\rm irr}\,\|}
  \equiv 
  \sqrt{h}\,L_\|^{\tp_\mu \tp_\nu}\,
  \biggl[\frac{\delta\Delta\Stot}{\delta\tp_\nu}
  \biggr]_{\mbox{\raisebox{3pt}{\scriptsize{$\|$}}}}\,,
\label{eq_tp_mu_para}\\
 \dot\tn - [\dot\tn]_{\rm rev} 
  &\equiv [\dot\tn]_{\rm irr} 
   \equiv \sqrt{h}\,L^{\tn \tn}\, \frac{\delta\Delta\Stot}{\delta \tn}\,.
\label{eq_tn}
\end{align}
We here do not consider the direct coupling 
between the variables $\varepsilon_\mu$ and $\tp_\mu$\,, 
expecting that it is negligible. 
Note also that $\tr\dot{\varepsilon}\equiv g^{\mu\nu}\,\dot{\varepsilon}_{\mu\nu}
=h^{\mu\nu}\,\dot{\varepsilon}_{\mu\nu}$ 
can be replaced by $(\tr\varepsilon)^\cdot$ at this order of approximation 
because $(\tr\varepsilon)^\cdot-\tr\dot{\varepsilon}
=(h^{\mu\nu})^\cdot\,\varepsilon_{\mu\nu}=-2N K^{\mu\nu}\,\varepsilon_{\mu\nu}$\,.
As we have done for fluids, we make the following irreducible decompositions 
under the group $\mathrm{O}(D)$ in a local inertial frame:
\begin{align}
  L^{\varepsilon_{\langle\mu\nu\rangle}\varepsilon_{\langle\rho\sigma\rangle}} 
  &\equiv  L^{\rm t}\,
  h_{\langle\mu}^{\mu'} h_{\nu\rangle}^{\nu'} h_{\mu'\rho}h_{\nu'\sigma} \,,\\
 L_\|^{\varepsilon_{\mu}\varepsilon_{\nu}} &\equiv L^{\rm v}\,h_{\mu\nu} \,,\\
 L^{\tr\varepsilon\,\tr\varepsilon} &\equiv L^{\rm s}_1\,, \\
 L^{\tr\varepsilon\,\theta} &\equiv L^{\rm s}_2\,, \\
 L^{\theta\,\theta} &\equiv L^{\rm s}_3\,, \\
 L_\bot^{\tp_\mu \tp_\nu} &\equiv L_\bot\,u_\mu u_\nu\,,\\
 L_\|^{\tp_\mu \tp_\nu} &\equiv L_\|\,h_{\mu\nu}\,,\\
 L^{\tn\tn}&\equiv M\,.
\end{align}
Thus, by assuming that the reversible evolution of $\tp_\mu$ and $\tn$ 
is expressed 
with reversible currents $\tau_\mu^{{\mathrm{(r)}}\,\nu}$ and $\nu_{\mathrm{(r)}}^\mu$
(whose explicit form will be given later for a simple case)
as
\begin{align}
 [\,\dot{\tp}_\mu]_{\rm rev} = -\sqrt{-g}\,\nabla_\nu \tau_\mu^{{\rm (r)}\,\nu}\,, \qquad 
 [\dot{\tn}]_{\rm rev} = -\sqrt{-g}\,\nabla_\mu \nu_{\rm (r)}^\mu \,,
\end{align}
Eqs.\ \eq{eq_tp_mu_perp}--\eq{eq_tn} become 
\begin{align}
 \bigl[\,\nabla_\mu \bigl(p^\nu u^\mu+\tau_{\rm (r)}^{\mu\nu}\bigr)\bigr]_\bot 
  &= -\,c_\bot\,L_\bot\,(-u^\nu u_\lambda)\,\nabla_\mu\,
  \bigl[\, \ell^{\rm t}_2\,\varepsilon^{\langle\mu\lambda\rangle}
  -(1/T)\,\ell^{\rm t}_3\,K^{\langle\mu\lambda\rangle} \nn\\
 &~~~~~~~~~
  + \bigl( \ell^{\rm s}_4\,\tr\varepsilon + \ell^{\rm s}_5\,\theta
  - (1/T)\,\ell^{\rm s}_6\,\tr K \bigr)\,h^{\mu\lambda}\bigr]\,,\\
 \bigl[\,\nabla_\mu \bigl(p^\nu u^\mu+\tau_{\rm (r)}^{\mu\nu}\bigr)\bigr]_\| 
  &= -\,c_\|\,L_\|\,h^{\nu}_{~\lambda}\,\nabla_\mu\,
  \bigl[\, \ell^{\rm t}_2\,\varepsilon^{\langle\mu\lambda\rangle}
  -(1/T)\,\ell^{\rm t}_3\,K^{\langle\mu\lambda\rangle} \nn\\
 &~~~~~~~~~
  + \bigl( \ell^{\rm s}_4\,\tr\varepsilon + \ell^{\rm s}_5\,\theta
  - (1/T)\,\ell^{\rm s}_6\,\tr K \bigr)\,h^{\mu\lambda}\bigr]\,,\\
 \nabla_\mu \bigl(n \,u^\mu + \nu_{\rm (r)}^\mu\bigr)
 &= -\,\sqrt{h}\,(-\partial^2\ts/\partial\tn^2)\,M\,\nabla_\mu
  \bigl[\,\ell^{\rm v}_2\,h^{\mu\nu}\,\varepsilon_\nu
  + \ell^{\rm v}_3\,h^{\mu\nu}\,\partial_\nu(-\mu/T)\bigr]\,.
\end{align}
Assuming again that $L_\tp\equiv c_\bot L_\bot= c_\| L_\|$ and 
$L_\tn\equiv\sqrt{h}\,(-\partial^2\ts/\partial\tn^2)\,M$ are constant, 
the above equations can be written in the form of current conservations: 
\begin{align}
  \nabla_\mu T^{\mu\nu}=0\,,\qquad \nabla_\mu n^\mu =0\,,
\label{current_conservation_viscoelastic}
\end{align}
where the conserved currents are given by 
\begin{align}
 T^{\mu\nu} \equiv e\,u^\mu u^\nu + \tau_{\rm (r)}^{\mu\nu}
  + \tau_{\rm (d)}^{\mu\nu}\,,\quad
 n^\mu \equiv nu^\mu + \nu^\mu_{\rm (r)} + \nu^\mu_{\rm (d)}\,,
\end{align}
with the dissipative currents
\begin{align}
 \tau_{\rm (d)}^{\mu\nu}
  &\equiv L_\tp\,\ell_2^{\rm t}\, \varepsilon^{\langle\mu\nu\rangle}
  - (1/T)\,L_\tp\,\ell_3^{\rm t}\, K^{\langle\mu\nu\rangle} 
    + L_\tp\,\bigl[\,\ell_4^{\rm s}\,\tr \varepsilon
    + \ell_5^{\rm s}\,\theta
    - (1/T)\,\ell_6^{\rm s}\,\tr K\bigr] h^{\mu\nu}\,,
\label{dissipative_tau}\\
 \nu^\mu_{\rm (d)} &\equiv L_\tn\,\bigl[\, \ell_2^{\rm v}\, h^{\mu\nu}\,\varepsilon_{\nu}
 + \ell_3^{\rm v}\, h^{\mu\nu}\,\partial_\nu (-\mu/T)\bigr]\,.
\label{dissipative_nu}
\end{align}

As for the $b^A$-type variables,  the equations of linear regression 
Eqs.\ \eq{eq_varepsilon_munu}--\eq{eq_scalar} 
can be represented in the following form:
\begin{align}
 \Lie_u\varepsilon_{\langle\mu\nu\rangle}
   =&\, N^{-1}[\dot{\varepsilon}_{\langle\mu\nu\rangle}]_{\rm rev}
   - L_{\rm t}\ell_1^{\rm t}\,\varepsilon_{\langle\mu\nu\rangle} 
   +(L_{\rm t}\ell_2^{\rm t}/T)\,K_{\langle\mu\nu\rangle}  \,, 
\label{rheology_1}\\
 (\Lie_u\varepsilon_\mu)_\bot 
  =&\, N^{-1} \bigl[\dot{\varepsilon}_\mu\bigr]_{{\rm rev}\,\bot}\,,
\label{rheology_2}\\
 (\Lie_u\varepsilon_\mu)_\| 
  =&\, N^{-1} \bigl[\dot{\varepsilon}_\mu\bigr]_{{\rm rev}\,\|} 
  - L^{\rm v}\, h_\mu^{~\nu}\,
   \bigl[\ell_1^{\rm v}\,\varepsilon_{\nu}
          +\ell_2^{\rm v}\, \partial_\nu (-\mu/T)\bigr] \,, 
\label{rheology_3}\\
 \Lie_u(\tr\varepsilon) 
  =&\, N^{-1}\bigl[(\tr\varepsilon)^\cdot\bigr]_{\rm rev}\nn\\
   &\,
  - (L^{\rm s}_1\,\ell^{\rm s}_1 + L^{\rm s}_2\,\ell^{\rm s}_2)\,\tr\varepsilon
  - (L^{\rm s}_1\,\ell^{\rm s}_2 + L^{\rm s}_2\,\ell^{\rm s}_3)\,\theta
  + (L^{\rm s}_1\,\ell^{\rm s}_4 + L^{\rm s}_2\,\ell^{\rm s}_5)\,
  \frac{1}{T}\tr K\,,
\label{rheology_4}\\
 \Lie_u\theta
  =&\, N^{-1} \bigl[\,\dot{\theta}\,\bigr]_{\rm rev} \nn\\
  &\,-\, (L^{\rm s}_2\,\ell^{\rm s}_1 + L^{\rm s}_3\,\ell^{\rm s}_2)\,\tr\varepsilon
  - (L^{\rm s}_2\,\ell^{\rm s}_2 + L^{\rm s}_3\,\ell^{\rm s}_3)\,\theta
  + (L^{\rm s}_2\,\ell^{\rm s}_4 + L^{\rm s}_3\,\ell^{\rm s}_5)\,
  \frac{1}{T}\tr K\,.
\label{rheology_5}
\end{align}
They give the generally covariant extension of 
the {\it rheology equations} introduced in \cite{Eckart:1948,afky}, 
and describe the dynamics of plastic deformations.

The current conservations 
\eq{current_conservation_viscoelastic}--\eq{dissipative_nu} 
and the rheology equations \eq{rheology_1}--\eq{rheology_5} 
totally determine the time evolution of the variables $p_\mu$\,,\ $n$\,,\ 
$\varepsilon_{\mu\nu}$\,,\ $\varepsilon_{\mu}$\,, and $\theta$\,,
and thus constitute the set of the fundamental equations 
that govern the dynamics of relativistic viscoelastic materials.
Further studies of relativistic viscoelastic materials 
are performed in our subsequent paper \cite{fs2}.

We finally make a comment that the explicit forms of 
$[\dot{\varepsilon}_{\langle\mu\nu\rangle}]_{\rm rev}$\,, 
$\bigl[\dot{\varepsilon}_\mu\bigr]_{\rm rev}$\,, 
$\bigl[(\tr\varepsilon)^\cdot\bigr]_{\rm rev}$\,, 
and 
$\bigl[\,\dot{\theta}\,\bigr]_{\rm rev}$ 
depend on the system under consideration.
For example, for the case when the reversible, isentropic evolution 
describes that of elastic materials, 
they are given by
\begin{align}
 &\tau_{\rm (r)}^{\mu\nu} 
  = -\,2{\mathcal G}\,\varepsilon^{\langle\mu\nu\rangle} 
     -{\mathcal K}\,\bigl(\tr\varepsilon-a\,\theta)\,h^{\mu\nu}\,,\quad
 \nu_{\rm (r)}^\mu = 0 \,, \\
 &N^{-1}\bigl[\dot{\varepsilon}_{\langle\mu\nu\rangle}\bigr]_{\rm rev}
 = K_{\langle\mu\nu\rangle}\,,\quad
 N^{-1}\bigl[(\tr\varepsilon)^\cdot\bigr]_{\rm rev} = \tr K \,,
\label{Kbar_0_1}\\
 &\bigl[\dot{\varepsilon}_{\mu}\bigr]_{\rm rev}
 =\bigl[\,\dot{\theta}\,\bigr]_{\rm rev}=0\,.
\label{Kbar_0_2}
\end{align}
Here the non-negative constants ${\mathcal G}$ and ${\mathcal K}$ 
represent the shear and bulk modulus, respectively,  
and the constant $a$ is proportional to the coefficient of thermal expansion.
Equations \eq{Kbar_0_1} and \eq{Kbar_0_2} show that
the intrinsic metric $\bar{h}_{\mu\nu}$ does not vary 
for reversible processes $\bigl(\bigl[\dot{\bar{h}}_{\mu\nu}\bigr]_{\rm rev}
=\bigl[2N\barK_{\mu\nu}\bigr]_{\rm rev}
=2 N K_{\mu\nu}-\bigl[\dot{\varepsilon}_{\mu\nu}\bigr]_{\rm rev}=0\bigr)$\,.
A few other examples of isentropic evolutions are also given 
in \cite{fs2}.

\section{Conclusion and discussions}

In this paper, we have developed an entropic formulation of 
relativistic continuum mechanics, 
which includes the standard relativistic fluid theory. 
The discussion is based on the linear nonequilibrium thermodynamics, 
and we have proposed a local functional $\Stot$ 
which represents the total entropy of nonequilibrium states 
and is to be maximized in the course of evolution. 
We have applied this framework to constructing 
a relativistic theory of viscoelastic materials.  
As is intensively studied in our subsequent paper \cite{fs2}, 
this theory can deal with a wide class of continuum materials, 
including as special cases elastic materials, 
Maxwell materials, Kelvin-Voigt materials, 
relativistic viscous fluids, 
and the so-called simplified Israel-Stewart fluids, 
and thus is expected to be the most universal description of 
single-component (not necessarily relativistic) continuum materials.

Our entropy functional $\Stot$ is a local functional.
Thus, once the coefficients $\ell^{AB}\,,\cdots\,,\ell^{\mu I,\,\nu J}$ 
in Eq.\ \eq{total_entropy} are determined for a given material, 
one can explicitly evaluate the difference of the entropy 
of any configuration from that in the global equilibrium, 
$\Delta\Stot=\Stot-\Stot_0$\,, 
as long as the configuration is not far from the global equilibrium; 
one only needs to measure the local values of thermodynamic variables 
(such as the values of temperature at each point) 
and put the obtained data into $\Delta\Stot$\,.

The next step would be to extend the current formalism
as one can deal with more complicated systems 
like multicomponent viscoelastic materials 
or systems with extra variables like liquid crystals. 
Such extension is actually straightforward and is under current investigation.

Another interesting direction would be to extract from our analysis 
the information on the holography of gravitation. 
In fact, one would need to assign local entropy 
to the metric tensor when the system is analyzed in generic frames 
(other than the Landau-Lifshitz frame)  
and when material particles have nonvanishing accelerations \cite{fs3}. 
If this point is carefully investigated within the framework of the Bogoliubov-Born-Green-Kirkwood-Yvon hierarchy, 
we would be able to find out the microscopic degrees of freedom 
which need to be introduced to describe general relativity at large spacetime scales. 
Investigation along this line is now in progress, 
and will be reported elsewhere.

\section*{Acknowledgments}
The authors thank Hikaru Kawai for useful discussions 
and Teiji Kunihiro for his nice lectures on relativistic fluid mechanics.   
This work was supported by the Grant-in-Aid for the Global COE program 
``The Next Generation of Physics, Spun from Universality and
Emergence" from the Ministry of Education, Culture, Sports, 
Science and Technology (MEXT) of Japan. 
This work was also supported by the Japan Society for the Promotion of Science 
(JSPS) (Grant No.\,21$\cdot$1105) and by MEXT (Grant No.\,19540288).

\appendix

\section{A derivation of the entropy functional}
\label{entropy_derivation}

In this appendix, we derive the entropy functional
\eq{total_entropy} in the flat background,
where the local entropy is assumed to take a simple form, 
$\ts(x)=\ts\bigl(\tc^I(x)\bigr)$\,.
In the following, we consider the special cases 
where $N=1$ and $h_{ij}$ are constant.

As can be seen from the equilibrium condition \eq{equilibrium_condition}, 
all intensive parameters $\beta_I$ are spatially constant 
in the equilibrium states.
We take the region $\Sigma_x[L_{\rm s}]$ to be a $D$-dimensional torus 
with the coordinates $x^i$ having the period $L_{\rm s}$\,.
We assume that finite size effects become irrelevant when $L_{\rm s}\gg\epsilon_{\rm s}$\,.
Then the equilibrium values in $\Sigma_x[L_{\rm s}]$
are equal to the mean values of $c^I(x)$ in $\Sigma_x[L_{\rm s}]$:
\begin{align}
 c^I_0 = \frac{C^I}{V} \quad \bigl( V=\sqrt{h}\,L_{\rm s}^D \bigr)\,,\quad
  \mbox{or}\quad\tc_0^I=\frac{C^I}{L_{\rm s}^D}\,,
\end{align}
where $C^I=\int \rmd^D\by\,\tc^I(t,\by)$ is the total charge 
for the region $\Sigma_x[L_{\rm s}]$\,.
We Fourier-expand  $\tc^I(t,\by)$ as
\begin{align}
 \tc^I(t,\by) = \frac{1}{V}\sum_{\bk} \Exp{\ii \bk\cdot \by}\tc^I_\bk(t)
            = \frac{C^I}{V}
            +\frac{1}{V}\sum_{\bk\neq {\boldsymbol 0}}
             \Exp{\ii \bk\cdot \by}\tc^I_\bk(t) \,,
\end{align}
where $k_i= 2\pi n_i/L_{\rm s}$ ($n_i\in \mathbb{Z}$)\,. 
The entropy functional
\begin{align}
 \Stot\bigl(t;\,\Sigma_x[L_{\rm s}]\bigr)
 =\int_{\Sigma_x[L_{\rm s}]} \ts\bigl(\tc^I(t,\by)\bigr)
\end{align}
can then be expanded as follows:
\begin{align}
 \Stot
 &= V\, s(\tc^I_0(t))
  + \frac{1}{V}\int_{\Sigma_x[L_{\rm s}]}\rmd^D\by\,
    \frac{\partial \ts}{\partial \tc^I}(\tc^I_0(t))
    \sum_{\bk\neq {\boldsymbol 0}} \Exp{\ii \bk\cdot \by}\tc^I_\bk(t)\nn\\
  &\quad\, + \frac{1}{2V^2}
    \int_{\Sigma_x[L_{\rm s}]}\rmd^D\by\,
      \frac{\partial^2\ts}{\partial\tc^I\partial\tc^J}(\tc^I_0(t))
     \sum_{\bk\neq {\boldsymbol 0}}\sum_{\bk'\neq {\boldsymbol 0}}
                  \Exp{\ii(\bk+\bk')\cdot\by}\,\tc^I_\bk(t)\,\tc^J_{\bk'}(t)\nn\\
  &= \Stot_0 + \frac{1}{2V (\sqrt{h})^2} s_{IJ}^0\, \sum_{\bk\neq {\boldsymbol 0}} \tc^I_\bk(t)\,\tc^J_{-\bk}(t)\,,
\label{S_tot-proof}
\end{align}
where $s^0_{IJ}=\sqrt{h}\,(\partial^2\ts/\partial\tc^I\partial\tc^J)\bigr|_0$\,, 
and we have used $\int\rmd^D\by\,
\sum_{\bk\neq {\boldsymbol 0}} \Exp{\ii \bk\cdot \by}\tc^I_\bk(t)=0$\,.

$\Stot-\Stot_0$ can be written as 
\begin{align}
 -\,\frac{1}{2V(\sqrt{h})^2}\,\sum_\bk \, f_{IJ}(\bk)\,\tc^I_\bk(t)\,\tc^J_{-\bk}(t) \,,
\end{align}
with $f_{IJ}(\bk)=-\,s^0_{IJ}\,(1-\delta_{\bk,{\boldsymbol 0}})$\,.
However, this nonanalytic form of $f_{IJ}(\bk)$ is not desirable 
because it may change the long-distance behavior 
in the analysis based on the derivative expansion. 
The most desirable is such a function that filters out low-$\bk$ modes 
analytically, 
and one can take as such the following function: 
\begin{align}
 f_{IJ}(\bk) = \ell^i_{~I,}{}^j_{~J}\,k_i\,k_j\,,
\end{align}
where $\ell^i_{~I,}{}^j_{~J}$ is a positive definite tensor 
and is symmetric under the exchange $(iI)\leftrightarrow(jJ)$\,.
The (mild) increase of $f_{IJ}(\bk)$ in the region 
$\lvert\bk\rvert \gtrsim 2\pi/L_{\rm s}$ 
does not cause a problem 
because $\tc^I_\bk(t)$ are assumed to decrease rapidly in the same region.
Then by using the equality 
$\partial_i(\partial \ts/\partial \tc^I)
= (1/\sqrt{h})\,s_{IJ}\,\partial_i \tc^J$\,,
the entropy functional can be written as
\begin{align}
 \Stot
  &= \Stot_0
  -\, \frac{1}{2V(\sqrt{h})^2}\,\sum_\bk \, \ell^i_{~I,}{}^j_{~J}\,k_i\,k_j\,
    \tc^I_\bk(t)\,\tc^J_{-\bk}(t) \nn\\
  &= \Stot_0
  -\, \frac{1}{2\sqrt{h}}\int_{\Sigma_x[L_{\rm s}]}\!\!\rmd^D\by \, 
  \ell^i_{~I,}{}^j_{~J} \,
  \partial_i \tc^I(t,\by)\,\partial_j \tc^J(t,\by) \nn\\
  &= \Stot_0
  - \frac{\sqrt{h}}{2} \int_{\Sigma_x[L_{\rm s}]}\!\! \rmd^D\by \, 
 \ell^{iI,\,jJ}\,\partial_i\Bigl(\frac{\partial \ts}{\partial \tc^{I}}\Bigr)\,
               \partial_j\Bigl(\frac{\partial \ts}{\partial \tc^{J}}\Bigr)  \,,
\end{align}
where $\ell^{iI,\,jJ} \equiv \ell^i_{~K,}{}^j_{~L}\,
(s_0^{-1})^{KI}\,(s_0^{-1})^{LJ}$\,.
For a general metric, $\Stot$ will have the following form:
\begin{align}
 \Stot
  = \Stot_0 - \frac{1}{2} \int_{\Sigma_x[L_{\rm s}]} \!\!\rmd^D \bx\,N\sqrt{h}\, 
  \ell^{\mu I,\nu J}\,\nabla_\mu\Bigl(\frac{\partial\ts}{\partial\tc^{I}}\Bigr)\,
                     \nabla_\nu\Bigl(\frac{\partial\ts}{\partial\tc^{J}}\Bigr) \,,
\end{align}
where the coefficients have only the spatial components, 
$\ell^{\mu I,\nu J}\,u_\nu=0$\,.
If there further exist additional $b^A$-type variables, 
this functional will have the more general form \eq{total_entropy}.

\baselineskip=0.85\normalbaselineskip


\begin{thebibliography}{9}
\setlength{\itemsep}{-2pt}

\bibitem{LL_stat}
  L.~D.~Landau and E.~M.~Lifshitz, 
  ``Statistical Physics, Part 1'', 
 Butterworth-Heinemann (1980).

\bibitem{Eckart:1940te}
  C.~Eckart,
  Phys.\ Rev.\  {\bf 58}, 919 (1940).

\bibitem{LL_fluid}
  L.~D.~Landau and E.~M.~Lifshitz, 
  ``Fluid Mechanics'', 
  Butterworth-Heinemann (1987).

\bibitem{review1}
  N.~Andersson and G.~L.~Comer,
  Living Rev.\ Rel.\  {\bf 10}, 1 (2007)
  [arXiv:gr-qc/0605010].

\bibitem{review2}
  P.~Romatschke,
  Int.\ J.\ Mod.\ Phys.\  E {\bf 19}, 1 (2010)
  [arXiv:0902.3663 [hep-ph]].

\bibitem{Onsager:I}
  L.~Onsager,
  Phys.\ Rev.\ {\bf 37}, 405 (1931).

\bibitem{Onsager:II}
  L.~Onsager,
  Phys.\ Rev.\ {\bf 38}, 2265 (1931).

\bibitem{Casimir}
  H.~B.~G.~Casimir,
  Rev.\ Mod.\ Phys.\  {\bf 17}, 343 (1945).

\bibitem{de_Groot-Mazur}
  S.~R.~de Groot and P.~Mazur, 
  ``Non-Equilibrium Thermodynamics'', 
  Dover (1984).

\bibitem{Eckart:1948}
  C.~Eckart,
  Phys.\ Rev.\  {\bf 73}, 373 (1948).

\bibitem{afky}
  T.~Azeyanagi, M.~Fukuma, H.~Kawai and K.~Yoshida,
  Phys.\ Lett.\  B {\bf 681}, 290 (2009)
  [arXiv:0907.0656 [hep-th]].

\bibitem{afky2}
  T.~Azeyanagi, M.~Fukuma, H.~Kawai and K.~Yoshida,
  to appear in the proceedings of Quantum Theory and Symmetries 6 (2010) 
  [arXiv:1004.3899 [hep-th]].

\bibitem{Arnowitt:1962hi}
  R.~L.~Arnowitt, S.~Deser and C.~W.~Misner,
  Gravitation: an introduction to current research, Louis Witten ed. (Wiley 1962), 
  chapter 7, pp 227-265, 
  [arXiv:gr-qc/0405109].

\bibitem{LL_elasticity}
  L.~D.~Landau and E.~M.~Lifshitz, 
  ``Theory of Elasticity'', 
  Butterworth-Heinemann (1986).

\bibitem{fs2}
  M.~Fukuma and Y.~Sakatani,
  ``Relativistic viscoelastic fluid mechanics,''
  Phys.\ Rev.\  E {\bf 84}, 026316 (2011)
  [arXiv:1104.1416 [cond-mat.stat-mech]].

\bibitem{fs3}
  M.~Fukuma and Y.~Sakatani,
  in preparation.

\end{thebibliography}
\end{document}